\documentclass{agujournal2019}
\usepackage{url} 
\usepackage[inline]{trackchanges} 
\usepackage{soul}

\usepackage[english]{babel}
\usepackage{amssymb}
\usepackage{amsmath}

\usepackage{algorithm}
\usepackage{algorithmic}

\journalname{}

\begin{document}

%
%

\title{Data assimilation for energy-aware hybrid models}

%
%




\authors{Igor Shevchenko\affil{1}~and Dan Crisan\affil{2}}


\affiliation{1}{National Oceanography Centre, European Way, Southampton, SO14 3ZH, UK}
\affiliation{2}{Department of Mathematics, Imperial College London, 180 Queen's Gate, London, SW7 2AZ, UK}




\correspondingauthor{Igor Shevchenko}{igor.shevchenko@noc.ac.uk}



\begin{keypoints}
\item 
Model fidelity sets a hard limit: Data Assimilation cannot fix what the model cannot represent -- better models enable better assimilation

\item 
Data Assimilation combined with a hybrid model achieves lower tracking error unlike Data Assimilation on a deficient model

\item 
Targeted assimilation in energetic regions matches full-domain accuracy, thus highlighting the importance of optimal observation design
\end{keypoints}

%
%

%
%


\begin{abstract}
This work integrates ensemble-based data assimilation (DA) with the energy-aware hybrid modeling
approach, applied to a three-layer quasi-geostrophic (QG) model of the Gulf Stream flow. 
Building on prior DA success in the QG channel regime, 
where stochastic corrections based on EOFs were effective, we show that this 
method fails to address persistent errors in the more complex, dynamically richer 
Gulf Stream setting.
To overcome this, we employ a hybrid model that controls energy at 
selected scales, maintaining dynamic consistency and physical realism. 
We evaluate the combined effect of hybrid modeling and DA, 
using a particle filter which combines model reduction, tempering, jittering, and nudging.

Numerical experiments show that the hybrid model 
reproduces both the large-scale jet and small-scale vortices seen 
in high-resolution reference simulations, but missing in the standard (non-hybrid) QG model. 
When  DA is incorporated, the hybrid model further reduces tracking error
and ensemble divergence.
Moreover, targeted assimilation from the most energetic 
region matches tracking error and uncertainty reduction of
full-domain networks, highlighting the critical importance of observation network design. 
These findings demonstrate that combining energy-aware hybrid modeling with 
ensemble-based DA enables high-fidelity, computationally efficient 
tracking of the reference solution even under sparse, noisy, localized observations. 
\end{abstract}

\section*{Plain Language Summary}
Scientists who study the ocean and atmosphere use computer models to help predict 
weather, climate, ocean currents, etc. These models are based on the laws of physics and 
allow us to simulate the behavior of complex systems.
To make these models run efficiently on the computer often requires simplifying some details, 
which can lead to missing important aspects of real-world behavior. 
To improve their accuracy and keep the models closer connected to reality, scientists combine 
model predictions with real-world observations, such as data from satellites, ships, 
or weather stations.

In our study, we developed and tested a new approach that combines two powerful ideas. 
First, we used a “hybrid” model that is specially designed to keep track of 
the ocean's total energy, even when running at a lower level of detail. 
This helps the model stay closer to the real ocean's behavior. Second, we used a method called 
“data assimilation,” which is a way of blending observations into the model as it runs. 
This keeps the simulation closely tied to what is really happening in the ocean, even when 
the data are limited or noisy.

We tested our method using a computer simulation of the Gulf Stream -- 
an important ocean current that affects weather patterns far beyond the Atlantic.
We found that our hybrid model 
could more accurately track the Gulf Stream than the standard non-hybrid model. 
Remarkably, we showed that it’s 
possible to get the same improvement in accuracy by focusing observations only on the most 
active part of the Gulf Stream as you would by observing the entire area in detail. 
This means that targeted measurements can be just as effective as widespread ones -- an 
important finding for designing future ocean monitoring systems.

%
%

%


%
%
%
%

\section{Introduction}
In recent years, Geophysical Fluid Dynamics (GFD) has seen a surge in the application 
of data-driven techniques, particularly those based on machine learning (ML), 
which encompasses a broad set of statistical and computational algorithms 
for pattern discovery and predictive 
modeling~\cite{mendez_ianiro_noack_brunton_2023,guillaumin2021machine}. 
Within ML, deep learning (DL), a family of neural network methods, has garnered significant 
attention for their potential to model nonlinear processes and handle large, 
complex datasets~\cite{reichstein2019deep,weyn2019can}. 
These data-driven approaches are increasingly explored as complements or alternatives to 
traditional physics-based modeling frameworks in weather and climate science.

Applications and advances of ML and DL techniques in GFD span
subgrid-scale parameterizations~\cite{brenowitz2018prognostic,gentine2018could,maulik2019subgrid},
short- to medium-term forecasting~\cite{weyn2019can,farchi2021using},
and computationally efficient emulation of expensive simulations~\cite{yuval2020stable,guillaumin2021machine}.
Despite their flexibility and computational advantages, data-driven methods encounter several 
persistent challenges:
limited availability of high-quality data, particularly for rare events~\cite{reichstein2019deep},
lack of guaranteed physical consistency~\cite{karpatne2017theory,beucler2021enforcing},
limited generalization to unseen regimes~\cite{farchi2021using,geer2021learning},
and poor interpretability, which makes it difficult 
to diagnose errors or build trust among domain scientists~\cite{guillaumin2021machine,yuval2020stable}.
These issues are especially acute in GFD, where the need for robust long-term predictions, 
reliable uncertainty quantification, and operational interpretability are paramount. 
As a result, there is growing consensus that purely data-driven models alone are unlikely to replace 
physics-based models in the foreseeable future~\cite{reichstein2019deep,guillaumin2021machine}.

The limitations of data-driven methods have motivated the development of hybrid modeling strategies 
that seek to integrate the strengths of data-driven methods (flexibility, efficiency, data utilization) with 
the reliability and interpretability of physical models~\cite{SB2022_J2,SC2024_J1,reichstein2019deep}. 
Recent work has explored physics-informed neural networks (PINNs)~\cite{karniadakis2021physics},
hybrid dynamical-ML systems~\cite{brenowitz2018prognostic,beucler2021enforcing},
ML models that enforce energy and mass conservation~\cite{beucler2021enforcing},
and energy-aware hybrid models~\cite{SC2024_J1}.

While energy-aware hybrid models have shown promising improvements over 
standard low-resolution GFD simulations, challenges remain. Notably, robust data assimilation 
(DA), i.e. the systematic integration of observational data into models, 
has often been treated separately from hybrid model development. 
Ensemble-based DA methods, including Ensemble Kalman Filters and particle filters, 
offer powerful tools for state estimation and uncertainty quantification in physics-based 
models~\cite{houtekamer2005ensemble,Potthast_et_al2019,vanleeuwen2019particle,CCHWS2019_1,CCHPS2020_J2}. 
Ensemble Kalman filters update an ensemble by a linear, covariance-based transformation and 
therefore work best when the forecast-observation operator is roughly linear and the errors are 
close to Gaussian. Particle filters, by contrast, approximate the full posterior with weighted particles and 
make no distributional assumption, but they may suffer weight collapse 
unless one uses huge ensembles, frequent resampling, and special techniques like jittering or tempering.

\noindent
{\bf Gap and novelty.}
The potential synergy between ensemble-based DA (operating in physical space) and hybrid 
modeling strategy operating in phase space -- the hyper-parameterization (HP) approach --
remains an unexplored frontier.
Most previous studies have either focused on improving hybrid models in isolation or 
advancing DA techniques for purely physics-based systems. Integrating these two paradigms offers 
the potential for both improved fidelity (by reducing tracking error and enhancing feature 
representation) and computational efficiency (through the use of lower-resolution, energy-aware 
hybrid models).
In this study, we address this gap by combining 
the hybrid HP approach (that enhances traditional low-resolution GFD models 
with energy-aware hybrid models) with 
an ensemble-based DA framework 
(specifically, a particle filter that incorporates model reduction, tempering, jittering, and 
nudging) to systematically assimilate observational data and further reduce model error.

The core novelty of our work lies in combining ensemble-based DA in physical space with the 
hybrid HP approach in phase space, creating a unified framework that leverages the strengths of 
both methodologies.
This allows us to achieve high accuracy and physical realism with significantly lower 
computational cost, compared to high-resolution physics-only models or data-driven methods alone.

Our approach is validated on a three-layer quasi-geostrophic (QG) model configured for 
beta-plane Gulf Stream flow -- a canonical benchmark for ocean modeling, 
e.g.~\cite{Karabasov_et_al2009,Ryzhov_etal_2019,Berloff_etal_2021,Sun_etal_2021}.
By rigorously comparing the hybrid DA approach with traditional models, 
we demonstrate marked improvements in tracking accuracy, flow feature representation, and 
overall predictive skill, even under conditions of sparse and noisy observational data.

By combining ensemble-based DA with the hybrid HP approach, 
we provide notable benefits and operational flexibility for oceanic and atmospheric modelers: 
low computational cost (enabled by low resolutions), 
high-fidelity results (via hybrid modeling), 
and the capacity to assimilate diverse observational datasets from 
drifters, satellites, etc.

\section{Hybrid hyper-parameterization models}
The hybrid approach assumes that a low-resolution GFD model 
may fail to capture the reference flow features not because of resolution limits, but 
due to insufficient energy. 
To recover these features, it
adjusts the energy distribution of the low-resolution solution to align it with that of the reference
by selectively modifying energy at specific spatial scales.

To explain a hybrid model, we start from the transport equation (which we call the reference model)
for a quantity~$\phi$\\
\begin{equation}
\partial_t\phi+\mathbf{v}\cdot\nabla\phi=\mathbf{F}(\phi)
\label{eq:pde} 
\end{equation}
with $\mathbf{v}(t,\cdot)$ being the velocity vector; the dot in the argument of $\mathbf{v}$ means the space dependence.
Equation~\eqref{eq:pde} is usually solved at high resolution, and every run is computationally prohibitive,
making long-term or large ensemble simulations infeasible.
A practical alternative is to employ a hybrid model that operates at lower resolution,
yet retains the ability to reproduce resolved on the coarse-grid flow features of the 
reference flow. The latter is defined 
as the projection of a high-resolution solution of equation~\eqref{eq:pde} onto the 
coarse grid used by the low-resolution hybrid model~\eqref{eq:hybrid}.
Before introducing the hybrid model, it is helpful to 
redefine $\phi$ and $\mathbf{v}$ as follows $\phi:=\mathcal{P}(\phi)$
and $\mathbf{v}:=\mathcal{P}(\mathbf{v})$, with $\mathcal{P}$ being a projector
from high to low resolution.

The hybrid model for equation~\eqref{eq:pde} reads as follows:\\
\begin{equation}
\partial_t\psi+(\mathbf{u}+\mathbf{A}(\mathbf{u},\mathbf{v}))\cdot\nabla\psi=\mathbf{F}(\psi)+\mathbf{G}(\psi,\phi),
\quad \mathbf{G}(\psi,\phi):=\eta(\mathbf{M}(\psi,\phi)-\psi)
\label{eq:hybrid} 
\end{equation}
with $\eta$ being the nudging strength, $\mathbf{u}$ is the velocity vector, and $\mathbf{M}$ is the multi-scale decompositions defined as\\
\begin{equation}
\mathbf{M}(\psi,\phi):=\sum\nolimits^S_{s=1}\lambda_s\mathcal{M}_s(\psi,\widehat{\phi}),\quad 
\widehat{\phi}:=\frac{1}{m}\sum\nolimits_{i=\mathcal{U}_I}\phi_i,
\label{eq:m} 
\end{equation}
where $\mathcal{M}_s  : \widehat{\phi}\rightarrow\widehat{\phi}_s$ is an operator extracting the $s$-scale flow dynamics, $\widehat{\phi}_s$, from $\widehat{\phi}$, which includes all 
(both underresolved and resolved on the computational grid) spatial scales,
$\lambda_s$ is the amplitude of $\widehat{\phi}_s$, $S$ is the total number of scales 
in the decomposition.
The neighborhood is a set of $m$ fields $\phi_i$ and
$\mathbf{v}_i$ ($i\in\mathcal{U}_I$) nearest in $l_2$-norm to the hybrid solution $\psi(t,\cdot)$;
$\mathcal{U}_I$ is a set of timesteps indexing the discrete reference solutions 
$\phi$ and $\mathbf{v}$ in the neighbourhood of $\psi(t,\cdot)$.
Note that equation~\eqref{eq:hybrid} is solved on a grid coarser than the one used in equation~\eqref{eq:pde}. 
Therefore, in order to use $\phi$ and $\mathbf{v}$ in equation~\eqref{eq:hybrid},
$\phi$ and $\mathbf{v}$ need to be projected from the fine to the coarse grid.

The velocity corrector is defined as\\
\begin{equation}
\mathbf{A}(\mathbf{u},\mathbf{v}):=\sum\nolimits^S_{s=1}\gamma_s\mathcal{M}_s(\widehat{\mathbf{v}}),\quad 
\widehat{\mathbf{v}}:=\frac{1}{m}\sum\nolimits_{i=\mathcal{U}_I}\mathbf{v}_i
\label{eq:a} 
\end{equation}
with index $i$ being the same as the one in~\eqref{eq:m},
and $\gamma_s$ is the amplitude of $\widehat{\mathbf{v}}_s$.

As an alternative, one can consider a stochastic velocity correction\\
\begin{equation}
\mathbf{A}(\mathbf{u},\mathbf{v}):=
{\color{blue}\sum\nolimits^S_{s=1}\gamma_s\mathcal{M}_s(\widehat{\mathbf{v}})dt}+\sum\nolimits^S_{s=1}\mathcal{M}_s(\widehat{\mathbf{v}})\circ dW^s_t,
\label{eq:dw} 
\end{equation}
with independent Brownian motions $dW^s_t$; we refer to the first term (highlighted in blue)
as ``v-nudging'' when we discuss particle filters.
In this case, the hybrid model~\eqref{eq:hybrid} becomes\\
\begin{equation}
d\psi+(\mathbf{u}\,dt+\mathbf{A}(\mathbf{u},\mathbf{v}))\cdot\nabla\psi=(\mathbf{F}(\psi)+\mathbf{G}(\psi,\phi))\,dt\,.
\label{eq:shybrid} 
\end{equation}
We refer to the nudging term $\mathbf{G}$ as G-nudging.
The velocity correction~\eqref{eq:dw} turns the hybrid model~\eqref{eq:shybrid} into a stochastic hybrid, which 
allows the application of the ensemble-based DA methodology explained in the next section.

{\bf Reference data acquisition}. 
To obtain the reference solution $\{\phi_{i \in [1,n]}\}$ and velocity field
$\{\mathbf{v}_{i \in [1,n]}\}$ (with $n$ being the number of records),
we simulate the reference model~\eqref{eq:pde} at high resolution and project 
the resulting $\phi$ and $\mathbf{v}$ fields onto the coarse grid used by the hybrid
model~\eqref{eq:hybrid}. It is important to note that the high-resolution 
simulation is performed only once, prior to the hybrid run, 
to generate the reference data. This reference simulation is typically much shorter 
than the subsequent hybrid model simulation.
If observational data are also available, they can also be projected onto the hybrid model 
grid and incorporated into the reference dataset. 
In this case, the combined reference dataset becomes 
$\{\phi_{i \in [1,n]}, \widetilde{\phi}_{j \in [1,n_o]}\}$ and
$\{\mathbf{v}_{i \in [1,n]}, \widetilde{\mathbf{v}}_{j \in [1,n_o]}\}$,
where the tilde indicates observed variables and $n_o$ is the number of observational records.
In scenarios where no numerical simulations are available, the hybrid model can 
operate solely based on observational data.

{\bf The multi-scale decomposition} controls the amount of energy injected into or 
extracted from the hybrid model at a given spatial scale. 
The operator $\mathcal{M}_s$ can be interpreted either as a spatial filter or 
as an evolutionary model that computes the $s$-scale component of the flow, 
denoted by $\widehat{\phi}_s$.
In this work, we employ spectral filtering (e.g.~\cite{Sagaut2006}), 
based on the discrete Fourier transform, to isolate flow dynamics at specific spatial scales. 
While spectral filtering is our method of choice, other filtering techniques may be used 
depending on the specific requirements of the problem.

{\bf The neighborhood} is defined as a set of $M$ fields 
$\phi_i$ and $\mathbf{v}_i$ ($i \in \mathcal{U}_I$) that are closest to the 
hybrid solution $\psi(t,\cdot)$ in the $l_2$-norm, within the reference phase 
space, i.e. the phase space associated with model~\eqref{eq:pde}.
In all simulations presented below, we fix $M = 10$, 
following the original HP method~\cite{SB2021_J1}. 
We emphasize that this parameter has limited impact on the results, 
as the optimization procedure is designed to compute amplitudes $\lambda_s$ and $\gamma_s$ 
that ensure the hybrid model’s energy remains within the prescribed reference energy band.
The choice of norm and the method used to construct the neighborhood
are not restricted to those applied in this study; alternative formulations may be employed 
depending on specific problem requirements. 
The approach used here is arguably the simplest and has proven 
sufficient for the objectives of this work.

\textbf{The number of spatial scales} $S$ is determined based on the resolution capacity of 
the hybrid model and the energy discrepancy between the reference and low-resolution solutions 
across different scales. A spatial scale is considered resolved if it spans at least 10 grid points,
a criterion motivated by the dispersion characteristics of the CABARET scheme~\cite{Karabasov_et_al2009}, which is used to solve the QG model described below.

Determining the \emph{optimal} multi-scale decomposition, i.e. the one that minimizes 
the discrepancy between the reference and hybrid solutions over a sampled time interval
would require a computationally expensive optimization procedure. 
Such a process, while feasible, lies beyond the scope of this work, and is of limited practical 
interest, since it is tailored specifically to the QG configuration used here.
More importantly, such an optimal decomposition is unnecessary for our purposes, as
our measure of goodness is satisfied as long as the hybrid solution remains within 
the reference energy band.
To compute a multi-scale decomposition that meets this criterion, we compare 
the energy spectra of the reference and low-resolution solutions. 
Spatial scales at which the energy difference is relatively large compared to other 
scales are selected for decomposition.
Since we employ spectral filtering in our scale decomposition, 
it is natural to use the energy spectral density $E_S$ as our diagnostic metric. 
Using kinetic and potential energy directly would require recalculating those quantities 
for each decomposed scale, thus making the spectral density a more practical and targeted choice.

{\bf Energy calculation}. To ensure the total energy of the hybrid model remains within the reference energy band, 
we compute both kinetic and potential energies. If potential energy data is unavailable, 
only the kinetic energy should be used. Notably, it is more efficient to precompute the 
energies for all reference data before running the hybrid model, thereby avoiding redundant 
energy calculations during the simulation.

{\bf Optimization method}.
The final component of the hybrid approach is the optimization method. 
In this study, we employ Powell's method~\cite{Powell1964}, a derivative-free 
optimization algorithm. Gradient-based methods are intentionally avoided, as they introduce 
unnecessary complexity for the objectives considered in this study.
While the nudging strength $\eta$ can, in principle, be treated as an additional optimization 
parameter, we keep it fixed at $\eta=0.02$ throughout this work as in~\cite{SC2024_J1}. 
Instead, the optimization is performed over the scale 
amplitudes $\lambda_s$ and $\gamma_s$, such that for all $t \ge 0$, one of the 
following energy-matching criteria is satisfied:\\
\begin{equation}
 \begin{aligned}
(C1) & & \min\limits_{\lambda_s,\gamma_s\ s\in[1,S]} \|\overline{E}(\phi)-E(\psi(t,\cdot))\|_2
 \end{aligned}
\label{eq:opt1}
\end{equation}
\begin{equation}
 \begin{aligned}
(C2) & & \min\limits_{\lambda_s,\gamma_s\ s\in[1,S]} \|E(\widehat{\phi})-E(\psi(t,\cdot))\|_2\,.
\end{aligned}
\label{eq:opt2}
\end{equation}
Here, $E(\psi(t,\cdot))$ denotes the total energy of the hybrid solution 
$\psi$, $\overline{E}(\phi)$ is the time-averaged total energy of 
the reference solution $\phi$, 
and $E(\widehat{\phi})$ is the total energy of $\widehat{\phi}$. 
The objective of the optimization is to align 
the hybrid solution energy $E(\psi(t,\cdot))$ with 
either $\overline{E}(\phi)$ or $E(\widehat{\phi})$, depending on the selected criterion.
The choice of optimization strategy plays a crucial role not only in the accuracy of 
the hybrid model but also in its computational efficiency. 
In our implementation, we apply optimization at the final time step of every 24-hour interval, 
with a simulation time step of 0.5 hour~\cite{SC2024_J1}. 

\section{Ensemble-based data assimilation methodology}
In this section we use the formalism of nonlinear filtering 
to discuss the ensemble-based data assimilation methodology adopted from~\cite{CCHWS2020_4,CCHWS2019_3}
for the use with the energy-aware hybrid model.
Consider a probability space $(\Omega,\mathcal{F},\mathbb{P})$, 
where we define two stochastic processes: the signal process $Z$ and 
the observation process $Y$. The signal process (also referred to as a reference or true state) is 
the solution of the reference model~\eqref{eq:pde} at high resolution 
projected onto the coarse grid, $G_s$, used in the hybrid model~\eqref{eq:hybrid};
we use a point-to-point projection, but other projections can be used instead, e.g.
spectral filters, spatial averaging, interpolation schemes, etc. 
The observation process $Y$ models observational data.

The  filtering problem is to estimate the posterior distribution of the signal $Z_t$, 
denoted by $\pi_t$, conditioned on the observations $Y_s$, $s\in [0,t]$. 
In the present setting, the observations correspond to noisy measurements of the reference state,
sampled at discrete times 
and
at designated locations (called weather stations) on a data grid $G_d$. 
Data assimilation is conducted at these observation times, referred to as 
assimilation times.

The simplest particle filter known as the bootstrap filter 
(see Section~\ref{sec:bootstrap} for further details) employs an ensemble of $N$ particles 
that evolve according to the signal dynamics between assimilation times. 
At each assimilation instance, particles are re-weighted based on the likelihood of their 
positions given the newly acquired observational data. A new ensemble is subsequently generated 
by resampling $N$ times (with replacement) from the weighted particles. 
This resampling mechanism favors particles with higher likelihoods (closer to the reference 
trajectory), which may be duplicated, while particles with lower likelihoods are discarded. 
As a consequence, the ensemble is expected to remain closer to the reference state 
than an ensemble evolving purely under the signal dynamics.

In our setting, implementing the bootstrap particle filter requires simulating multiple 
instances of the signal, which involves solving the reference model~\eqref{eq:pde} on a high 
resolution grid $G_f$. 
This is computationally intensive, and given that data assimilation is performed over 
thousands of steps, the total computational cost becomes prohibitive.
To address this challenge, we substitute the reference signal with a computationally 
cheaper proxy. Specifically, we introduce a process $X$, defined on the same probability space, 
whose sample paths are more efficient to simulate. In our case, $X$ is the solution 
to the stochastic hybrid model~\eqref{eq:shybrid}, evaluated on the coarser signal grid $G_s$, 
which significantly reduces the runtime of each simulation. 
This substitution constitutes a model reduction strategy, wherein the state space is 
reduced from $G_f$ to $G_s$. Such model reduction is not only crucial for the feasibility of 
our DA approach but is also justified rigorously.
The posterior distribution $\pi_t$ depends continuously on 
the prior distribution of the signal and the observational data. 
Consequently, replacing the reference signal distribution with a proxy distribution 
yields a reliable approximation of $\pi_t$, provided that the proxy is sufficiently close 
to the original in a suitably chosen topology on the space of probability measures.
The proximity between the original and proxy distributions is ensured
by controlling energy at specified spatial scales in the hybrid model.
This is achieved using the multi-scale decomposition and the optimization 
procedure explained earlier. It is important to note that our objective is not to approximate 
the reference state pathwise, but rather to approximate its posterior distribution.

In this work, the reference state is deterministic, whereas the process $X$ is stochastic. 
As demonstrated in~\cite{CCHWS2019_1,CCHPS2020_J2} the distribution of $X$ can be visualized 
using an ensemble of particle trajectories, each of which is a solution of the 
stochastic hybrid model on the grid $G_s$, driven by independent families of Brownian motions. 
Within the framework of uncertainty quantification, the discrepancy between the 
distributions of the reference state and $X$ is interpreted as model uncertainty. 
Typically, the ensemble forms a spread centered around the reference trajectory, 
with the magnitude of this spread reflecting the level of uncertainty.
This uncertainty can be visualized by comparing projections of the reference trajectory and 
those of the particle ensemble at selected grid points. 
Naturally, increasing the resolution of the grid $G_s$ leads to a reduction in the 
discrepancy between the two distributions, thus reducing the spread. 
However, this refinement comes at the cost of increased computational burden for 
generating particle trajectories. One of the central objectives of data assimilation is 
to reduce this spread (i.e. the uncertainty) without resorting to grid refinement.

The mean of the particle ensemble produced by the DA procedure, $\widehat{Z}_t$, 
serves as a pointwise estimator of the reference state $Z_t$, while
the ensemble spread provides a measure of the approximation error $Z_t - \widehat{Z}_t$. 
The DA method employed here is asymptotically consistent: as $N \to \infty$, 
the empirical distribution of the particles converges to the posterior distribution 
$\pi_t$~\cite{Crisan2002ASO}. Consequently, $\widehat{Z}_t$ converges to the conditional 
expectation of $Z_t$ given the observations, and the empirical covariance converges 
to the conditional expectation of $(Z_t - \widehat{Z}_t)(Z_t - \widehat{Z}_t)^T$ conditioned on 
the same observations.\footnote{$(Z_t - \widehat{Z}_t)^T$ denotes the transpose of $(Z_t - \widehat{Z}_t)$.}
This holds true under the assumption that particles evolve under the reference signal 
distribution. In our case, however, we employ a proxy distribution, and thus the 
limiting empirical distribution approximates $\pi_t$. The quality of this approximation is 
controlled by the resolution of the signal grid $G_s$.

The DA methodology presented below integrates 
the bootstrap particle filter with three additional methods: nudging, tempering, and jittering. 
The bootstrap particle filter, when used in isolation, fails to provide a reliable 
approximation of the posterior distribution. This is due to the fact that particle likelihoods 
tend to exhibit large variability, as individual particles quickly diverge from the reference 
state in different directions. This divergence, when compared against observational data, 
results in a situation where one or a few particles dominate the likelihoods, 
leading to repeated selection of only those particles during resampling. 
Such degeneracy undermines the representativeness of the ensemble.
The inclusion of nudging, tempering, and jittering mitigates this effect 
by promoting a more evenly spread ensemble. As we elaborate below, these techniques ensure 
that the particle set better captures the underlying posterior distribution.

As outlined earlier, we employ an ensemble of solutions (called particles) of the 
stochastic hybrid model~\eqref{eq:shybrid}, each driven by an independent realization of Brownian noise $W$. 
The number of independent Brownian motions is set to the number of scales used in the multi-scale
decomposition, namely $S=2$ as in our previous work~\cite{SC2024_J1}.

In our setting, the observational data $Y_t$ is modeled as an $M$-dimensional 
stochastic process comprising noisy measurements of the velocity field $\mathbf{u}$, 
recorded at specific locations on the data grid $G_d$. Formally, we define

$$
Y_t:={\rm P}^s_d(Z_t)+\chi,
$$
where ${\rm P}^s_d: G_s \rightarrow G_d$ denotes a projection operator from the signal grid $G_s$ to the data grid $G_d$, and 
$\chi$ is a random noise vector distributed as $\mathcal{N}(\mathbf{0}, I_\sigma)$. 
Here, $\mathbf{0} = (0,\ldots,0)$ is the mean vector and 
$I_\sigma = \operatorname{diag}(\sigma_1^2, \ldots, \sigma_M^2)$ 
is a diagonal covariance matrix representing the measurement uncertainty 
at each observation point.
Rather than prescribing the standard deviation 
vector $\sigma = (\sigma_1, \ldots, \sigma_M)$ arbitrarily, 
we define it based on the empirical standard deviation of the velocity field over 
each coarse grid cell of the signal grid $G_s$. 
This choice reflects the conceptual interpretation of the coarse model 
as resolving only the spatially averaged dynamics at the scale of $G_s$. 
Since the observations are extracted as pointwise values from the reference solution, 
we treat the observational noise as capturing the discrepancy between these pointwise 
measurements and the corresponding local spatial averages.
Assuming that small-scale fluctuations are sufficiently ergodic, 
this justifies to model the point value as a cell average plus
a random fluctuation. We consider this fluctuation
to be the principal source of observational error in the DA framework.

We introduce the likelihood-weight function

\begin{equation}
 \mathcal{W}(\mathbf{X},\mathbf{Y})=
 \exp\left(-\frac12\sum\limits^M_{i=1}\left\|\frac{{\rm P}^s_d (X_i)-Y_i}{\sigma_i}\right\|^2_2\right)\,,
 \label{eq:prob_function}
\end{equation}
where $M$ denotes the number of observation points (weather stations). 
To measure the variability of particle weights~\eqref{eq:prob_function} we compute 
the effective sample size:

\begin{equation}
 {\rm ESS}(\overline{\mathbf{w}})=\left(\sum\limits^N_{n=1}\left(\overline{w}_n\right)^2\right)^{-1},\quad
 \overline{\mathbf{w}}:=\mathbf{w}\left(\sum\limits^N_{n=1}w_n\right)^{-1},\quad w_n:=\mathcal{W}(\mathbf{X}^{(n)},\mathbf{Y}^{(n)})\,,
 \label{eq:ess}
\end{equation}
where $N$ is the total number of particles.
The ESS takes values close to $N$ when the weights are approximately uniform, 
indicating a well-balanced ensemble. In contrast, it approaches 1 when the ensemble 
degenerates, i.e. when a small number of particles dominate the weights and the 
remaining particles contribute negligibly.
To maintain ensemble diversity, resampling is performed whenever the effective sample size 
drops below a prescribed threshold $N^*$. In the present work, we choose $N^* = 80$.

It is important to remark that the stochastic velocity corrector~\eqref{eq:dw} 
necessitates to use the v-nudging term. However, in what follows
we set this term to zero to avoid unnecessary complications in further explanations
of the particle filter. We will return to this term when explaining how to bridge 
the hybrid model and particle filters.

Another important aspect of this work is that the original DA methodology~\cite{CCHWS2020_4,CCHWS2019_3} 
does not use the G-nudging term in~\eqref{eq:hybrid}, i.e. $\mathbf{G}=0$, and
utilizes a set of specially-calibrated Empirical Orthogonal Functions (EOFs), $\xi_s$, in place of 
$\mathcal{M}_s(\widehat{\mathbf{v}})$ within the stochastic velocity corrector~\eqref{eq:dw}.
The corrector is therefore expressed as

\begin{equation}
\widetilde{\mathbf{A}}:=
{\color{blue}\sum\nolimits^S_{s=1}\gamma_s\xi_s dt}+\sum\nolimits^S_{s=1}\xi_s\circ dW^s_t,
\label{eq:dw2} 
\end{equation}
with $S$ and $\gamma$ now being the number of EOFs and their amplitudes, respectively.

In a nutshell, the calibration procedure for the original DA method involves
computing EOFs for the difference between passive Lagrangian particles advected by the velocity $\mathbf{v}$ 
at high-resolution ($513\times513$) and low-resolution ($129\times129$). 
For a detailed discussion of this calibration process, 
we refer the reader to~\cite{CCHWS2019_1,CCHPS2020_J2}.
When using the original DA method, it is necessary to replace
$\mathbf{A}$ with $\widetilde{\mathbf{A}}$
and $\mathcal{M}_s(\widehat{\mathbf{v}})$ with $\xi_s$, and also set $\mathbf{G}=0$
in all subsequent DA algorithms. Additionally, the energy constraint 
should be excluded from the minimization problems~\eqref{eq:opt11} and~\eqref{eq:opt21}, 
and only the cost function $\Phi(\gamma)$ should be minimized.

\subsection{Bootstrap Particle Filter \label{sec:bootstrap}}
In this section, we consider the most basic particle filtering method, 
known as the \textit{bootstrap particle filter}, also referred 
to as the \textit{Sampling Importance Resampling} filter~\cite{DFG2001} -- Algorithm~\ref{alg:bootstrap}.
The filter operates as follows:

\begin{algorithm}
\caption{\bf Bootstrap particle filter}
\begin{algorithmic}
\label{alg:bootstrap}
\FOR{$j=0,1,2,\ldots$}
\STATE{{\bf Solve} $d\psi^{(n)}+(\mathbf{u}^{(n)}\,dt+\mathbf{A}(\mathbf{u}^{(n)},\mathbf{v}))\cdot\nabla\psi^{(n)}=(\mathbf{F}(\psi^{(n)})+\mathbf{G}(\psi^{(n)},\phi))\,dt$}
\STATE{\qquad $t\in[t_j,t_{j+1}],\quad n\in[1,N].$}
\STATE{{\bf Receive observations} $\mathbf{Y}_{t_{j+1}}$}
\STATE{{\bf Compute} $\overline{\mathbf{w}}$ (\,$w_n:=\mathcal{W}(\psi^{(n)}_{t_{j+1}},\mathbf{Y}^{(n)}_{t_{j+1}}),\, n\in[1,N]$\,).}
\IF{${\rm ESS(\overline{\mathbf{w}})}<N^*$}
\STATE{$\psi_{t_{j+1}}:={\bf Resample}(\overline{\mathbf{w}})$}
\ENDIF
\ENDFOR
\end{algorithmic}
\end{algorithm}

Starting from an initial distribution of particles, each particle is propagated forward 
in time according to the stochastic hybrid model. 
Based on partial observations, $\mathbf{Y}_{t_{j+1}}$, of the reference state, 
the weights of the new particles are computed. If the effective sample size falls below a 
critical threshold $N^*$, the particles are resampled to eliminate those with small weights.

For high-dimensional problems such as the one considered in this study, the effective sample 
size tends to decrease rapidly, often collapsing to 1 due to sample degeneracy. 
This happens because particles diverge quickly from the reference state -- a discrepancy 
captured by the observation data (given that the measurement noise is not large in 
our case, i.e., the observations are relatively accurate). To counteract this, either 
resampling must be performed unpractically frequently, or a substantially larger number of 
particles must be used.

To maintain ensemble diversity, we instead employ three additional procedures: 
\textit{the tempering technique}, \textit{jittering} based on the 
\textit{Metropolis-Hastings Markov chain Monte Carlo (MCMC)} method, and \textit{nudging}. 
Each of these techniques is explained in the subsections that follow.

\subsection{Tempering and jittering \label{sec:tempering}}
We briefly outline the usage of \textit{tempering} and \textit{jittering}, referencing \cite{CCHWS2019_3} for further details. The core idea behind \textit{tempering} is to artificially flatten the particle weights by rescaling the log-likelihoods using a factor $\phi \in (0,1]$, referred to as the temperature. After this rescaling, resampling can be performed. This adjustment leads to a more diverse ensemble, as the effective sample size (ESS) becomes more balanced -- the temperature is chosen precisely to achieve this effect. Nevertheless, some particles may still be duplicated, necessitating the use of \textit{jittering} to restore ensemble diversity.

\textit{Jittering} serves to enhance ensemble diversity by replacing duplicated particles with newly generated ones. There are several strategies for introducing diversity. A straightforward method is to add random perturbations to particles. However, doing so may produce particles that no longer satisfy the stochastic hybrid model, potentially resulting in nonphysical model behavior. To mitigate this, we generate new particles by solving the stochastic hybrid model driven by a modified Brownian motion,
$\rho W + \sqrt{1 - \rho^2},d\widetilde{W}$, where $W$ is the original Brownian motion and $\widetilde{W}$ is an independent copy. The initial conditions for this modified model are identical to those in~\eqref{eq:shybrid}.

The perturbation parameter $\rho$ is chosen to balance proximity to the original particle positions with sufficient deviation to ensure ensemble diversity. In our experiments, we set $\rho = 0.9999$. Each new particle proposal is then subject to a Metropolis-Hastings acceptance step, with $M_1$ denoting the number of iterations; we use $M_1 = 20$. This ensures that the perturbations do not alter the target distribution. However, the choice of a fixed number of jittering steps may not be optimal. 
Further work is needed to investigate how to adapt this in relation to the perturbation parameter $\rho$.

Following the completion of the initial tempering-jittering cycle, the resulting ensemble represents an altered distribution rather than the target posterior. Thus, the procedure must be repeated: a new temperature value within the interval $(\varphi, 1]$ is selected to maintain a reasonable ESS, and the tempering-jittering steps are applied again. This cycle continues until the temperature reaches 1.0, thereby recovering the original target distribution. The full methodology is formalized in Algorithm~\ref{alg:temp_mcmc} below.

\begin{algorithm}
\caption{\bf Particle Filter with Tempering and MCMC}
\begin{algorithmic}
\label{alg:temp_mcmc}
\FOR{$j=0,1,2,\ldots$}

\STATE{{\bf Solve} $d\psi^{(n)}+(\mathbf{u}^{(n)}\,dt+\mathbf{A}(\mathbf{u}^{(n)},\mathbf{v}))\cdot\nabla\psi^{(n)}=(\mathbf{F}(\psi^{(n)})+\mathbf{G}(\psi^{(n)},\phi))\,dt$}
\STATE{\qquad $t\in[t_j,t_{j+1}],\quad n\in[1,N].$}
\STATE{{\bf Receive observations} $\mathbf{Y}_{t_{j+1}}$}
\STATE{{\bf Compute} $\overline{\mathbf{w}}$ (\,$w_n:=\mathcal{W}(\psi^{(n)}_{t_{j+1}},\mathbf{Y}^{(n)}_{t_{j+1}}),\, n\in[1,N]$\,).}
\IF{${\rm ESS(\overline{\mathbf{w}})}<N^*$}
\STATE{{\bf Find} $p$ such that $ESS(\overline{\mathbf{w}})\ge N^*$, where
$\overline{\mathbf{w}}$ is computed with}
\STATE{\qquad $w_n:=\mathcal{W}^{1/p}(\psi^{(n)}_{t_{j+1}},\mathbf{Y}^{(n)}_{t_{j+1}}),\, n\in[1,N]$.}
\FOR{$k=1:p$}
\STATE{{\bf Compute} $\overline{\mathbf{w}}$ (\, $w_n:=\mathcal{W}^{\phi_k}(\psi^{(n)}_{t_{j+1}},\mathbf{Y}^{(n)}_{t_{j+1}}),\, n\in[1,N]\,)$,
$\phi_k:=\frac{k}{p}$ \,}
\STATE{$\psi_{t_{j+1}}:={\bf Resample}(\overline{\mathbf{w}})$}
\FOR{$m=1:M_1$}

\STATE{${\color{black} \Xi^{(n)}:=
\sum\limits^S_{s=1}\mathcal{M}_s(\widehat{\mathbf{v}}) \circ \left(\rho\, dW^{(n)}_{s}+\sqrt{1-\rho^2}\,d\widetilde{W}^{(n)}_{s}\right)},\quad n\in[1,N]$}

\STATE{$\mathbf{A}(\widetilde{\mathbf{u}}^{(n)},\mathbf{v}):=
\sum\limits^S_{s=1}\gamma_s\mathcal{M}_s(\widehat{\mathbf{v}})dt+{\color{black}\Xi^{(n)}},\quad n\in[1,N]$}

\STATE{{\bf Solve} $d\widetilde{\psi}^{(n)}+(\widetilde{\mathbf{u}}^{(n)}\,dt+
\mathbf{A}(\widetilde{\mathbf{u}}^{(n)},\mathbf{v}))\cdot\nabla\widetilde{\psi}^{(n)}=(\mathbf{F}(\widetilde{\psi}^{(n)})+\mathbf{G}(\widetilde{\psi}^{(n)},\phi))\,dt$}
\STATE{\qquad $t\in[t_j,t_{j+1}],\quad n\in[1,N].$}

\FOR{n=1:N}

\STATE{$\alpha:=\left(\mathcal{W}(\widetilde{\psi}^{(n)}_{t_{j+1}},Y^{(n)}_{t_{j+1}})/\mathcal{W}(\psi^{(n)}_{t_{j+1}},Y^{(n)}_{t_{j+1}})\right)^{\phi_k}$}
\IF{$1\le\alpha$}
\STATE{$\psi^{(n)}_{t_{j+1}}:=\widetilde{\psi}^{(n)}_{t_{j+1}}$}
\ELSIF{$\mathcal{U}[0,1]<\alpha$}
\STATE{$\psi^{(n)}_{t_{j+1}}:=\widetilde{\psi}^{(n)}_{t_{j+1}}$}
\ENDIF
\ENDFOR
\ENDFOR
\ENDFOR
\ENDIF
\ENDFOR
\end{algorithmic}
\end{algorithm}

\newpage
\subsection{Nudging \label{sec:nudging}}
\textit{Tempering} combined with \textit{jittering} forms a powerful framework that allows for the proper reduction of stochastic spread in the presence of informative data, while simultaneously preserving ensemble diversity over extended time intervals. The effectiveness of this combined strategy 
hinges on the quality of the initial particle proposals. These proposals are generated by evolving the particles under 
the stochastic hybrid model (representing the proxy distribution) rather 
than the reference target distribution. To reduce the discrepancy introduced 
by this approximation, one can apply \textit{nudging}.

In this work we operate with two different nudging terms: v-nudging and G-nudging.
\textit{If the hybrid model is used with no particle filter}, then the 
v-nudging and G-nudging term enter 
the optimization criteria~\eqref{eq:opt1} and~\eqref{eq:opt2} with no changes;
in this case, they both serve the same purpose -- to keep the energy of the hybrid model
within the reference energy band.
\textit{If the hybrid model is used with the particle filter},
then the G-nudging is utilized as before, while the v-nudging term plays a double role --
it ensures the energy of the hybrid model lies within the reference energy 
band (as before) and also adjusts the solution of the stochastic hybrid model~\eqref{eq:shybrid} 
to keep the particle trajectories in the neighborhood of the reference system state. 

Thus, the hybrid stochastic solution $\psi$ becomes 
parameterized by $\gamma$, and the particle trajectories become depending on this parameter.
According to Girsanov's theorem, the new weights of these particles are given by

\begin{equation}
 \mathcal{W}({\psi}(\gamma),\mathbf{Y},\gamma)=
 \exp\left(-\Phi(\gamma)\right),
 \label{eq:girsanov}
\end{equation}

\noindent
where

$$
\Phi(\gamma):=\frac12\sum\limits_{i=1}^M\left\|\frac{P^s_d(\psi_{t_{j+1}}(\gamma))-\mathbf{Y}_{t_{j+1}}}{\sigma_i}\right\|^2_2+
{\color{blue}\int_{t_{j}}^{t_{j+1}}\left(\gamma^2_s\frac{ dt}{2}-\gamma_s dW_s\right)}.
$$

These weights quantify the likelihood of a particle’s position given the observations. 
The final (blue) term accounts for the change of measure from the reference distribution $\psi$ 
to the modified distribution $\psi(\gamma)$. Thus, it is natural to aim for values of 
$\gamma$ that maximize this likelihood. Equivalently, we can solve one of the following minimization problems:

\begin{equation}
 \begin{aligned}
(C1^*) & & \min\limits_{\lambda_s,\gamma_s\ s\in[1,S]} 
\Big[ \|\overline{E}(\phi)-E(\psi(t,\cdot))\|_2+\Phi(\gamma) \Big]
 \end{aligned}
\label{eq:opt11}
\end{equation}
\begin{equation}
 \begin{aligned}
(C2^*) & & \min\limits_{\lambda_s,\gamma_s\ s\in[1,S]} 
\Big[ \|E(\widehat{\phi})-E(\psi(t,\cdot))\|_2 + \Phi(\gamma) \Big]
\end{aligned}
\label{eq:opt21}
\end{equation}
in conjunction with the stochastic hybrid model~\eqref{eq:shybrid}.
These are combined criteria, which serve to maintain the hybrid model energy within
the reference energy band while ensuring the particle trajectories remain in the neighborhood of the
reference state; in this work we use criterion~\eqref{eq:opt21}, as 
criterion~\eqref{eq:opt2} for the hybrid model shows better results compared 
with criterion~\eqref{eq:opt1}~\cite{SC2024_J1}.
This is the approach adopted in this work. Thus, the DA
method with nudging is given by Algorithm~\ref{alg:nudging}.

\begin{algorithm}
\caption{\bf Particle Filter with Tempering, MCMC, and Nudging}
\begin{algorithmic}
\label{alg:nudging}
\FOR{$j=0,1,2,\ldots$}

\STATE{{\bf Solve} $d\psi^{(n)}+(\mathbf{u}^{(n)}\,dt+\mathbf{A}(\mathbf{u}^{(n)},\mathbf{v}))\cdot\nabla\psi^{(n)}=(\mathbf{F}(\psi^{(n)})+\mathbf{G}(\psi^{(n)},\phi))\,dt$}
\STATE{\qquad $t\in[t_j,t_{j+1}],\quad n\in[1,N].$}
\STATE{{\bf Receive observations} $\mathbf{Y}_{t_{j+1}}$}
\STATE{\color{black}{\bf Solve} $\min\limits_{\lambda_s,\gamma_s\ s\in[1,S]} 
\left[ \|E(\widehat{\phi})-E(\psi(t,\cdot))\|_2 + \Phi(\gamma) \right]$}
\STATE{{\bf Compute} $\overline{\mathbf{w}}$ (\,$w_n:=\mathcal{W}(\psi^{(n)}_{t_{j+1}},\mathbf{Y}^{(n)}_{t_{j+1}},\gamma),\, n\in[1,N]$\,).}
\IF{${\rm ESS(\overline{\mathbf{w}})}<N^*$}
\STATE{{\bf Find} $p$ such that $ESS(\overline{\mathbf{w}})\ge N^*$, where
$\overline{\mathbf{w}}$ is computed with}
\STATE{\qquad $w_n:=\mathcal{W}^{1/p}(\psi^{(n)}_{t_{j+1}},\mathbf{Y}^{(n)}_{t_{j+1}},\gamma),\, n\in[1,N]$.}
\FOR{$k=1:p$}
\STATE{{\bf Compute} $\overline{\mathbf{w}}$ (\, $w_n:=\mathcal{W}^{\phi_k}(\psi^{(n)}_{t_{j+1}},\mathbf{Y}^{(n)}_{t_{j+1}},\gamma),\, n\in[1,N]\,)$,
$\phi_k:=\frac{k}{p}$ \,}
\STATE{$\psi_{t_{j+1}}:={\bf Resample}(\overline{\mathbf{w}})$}
\FOR{$m=1:M_1$}

\STATE{${\color{black} \Xi^{(n)}:=
\sum\limits^S_{s=1}\mathcal{M}_s(\widehat{\mathbf{v}}) \circ \left(\rho\, dW^{(n)}_{s}+\sqrt{1-\rho^2}\,d\widetilde{W}^{(n)}_{s}\right)},\quad n\in[1,N]$}

\STATE{$\mathbf{A}(\widetilde{\mathbf{u}}^{(n)},\mathbf{v}):=
\sum\limits^S_{s=1}\gamma_s\mathcal{M}_s(\widehat{\mathbf{v}})dt+{\color{black}\Xi^{(n)}},\quad n\in[1,N]$}

\STATE{{\bf Solve} $d\widetilde{\psi}^{(n)}+(\widetilde{\mathbf{u}}^{(n)}\,dt+
\mathbf{A}(\widetilde{\mathbf{u}}^{(n)},\mathbf{v}))\cdot\nabla\widetilde{\psi}^{(n)}=(\mathbf{F}(\widetilde{\psi}^{(n)})+\mathbf{G}(\widetilde{\psi}^{(n)},\phi))\,dt$}
\STATE{\qquad $t\in[t_j,t_{j+1}],\quad n\in[1,N].$}

\FOR{n=1:N}

\STATE{$\alpha:=\left(\mathcal{W}(\widetilde{\psi}^{(n)}_{t_{j+1}},Y^{(n)}_{t_{j+1}},\gamma)/\mathcal{W}(\psi^{(n)}_{t_{j+1}},Y^{(n)}_{t_{j+1}},\gamma)\right)^{\phi_k}$}
\IF{$1\le\alpha$}
\STATE{$\psi^{(n)}_{t_{j+1}}:=\widetilde{\psi}^{(n)}_{t_{j+1}}$}
\ELSIF{$\mathcal{U}[0,1]<\alpha$}
\STATE{$\psi^{(n)}_{t_{j+1}}:=\widetilde{\psi}^{(n)}_{t_{j+1}}$}
\ENDIF
\ENDFOR
\ENDFOR
\ENDFOR
\ENDIF
\ENDFOR
\end{algorithmic}
\end{algorithm}

Note that, for the sake of simplicity, we omitted the explanation of how to
combine the minimization procedure with the Bootstrap filter and the tempering algorithm.
To implement this combination, one should apply the minimization step after receiving 
the observations $\mathbf{Y}_{t_{j+1}}$, 
as illustrated in Algorithm~\ref{alg:nudging}.

\section{Multilayer quasi-geostrophic model\label{sec:qg}}
In this section we apply the hybrid approach to the three-layer quasi-geostrophic (QG) 
model that describes the evolution of potential vorticity (PV) 
anomaly $\mathbf{q}=(q_1,q_2,q_3)$ (e.g.~\cite{Pedlosky1987}):\\
\begin{equation}
\partial_t q_j+\mathbf{v}_j\cdot\nabla q_j=F,\quad F:=\delta_{1j}F_{\rm w}-\delta_{j3}\,\mu\nabla^2\psi_j+\nu\nabla^4\psi_j-\beta\psi_{jx}, \quad j=1,2,3\, ,
\label{eq:pve}
\end{equation}
with $\boldsymbol{\psi}=(\psi_1,\psi_2,\psi_3)$ is the velocity streamfunction, $\delta_{ij}$ is the Kronecker symbol,
The planetary vorticity gradient is $\beta=2\times10^{-11}\, {\rm m^{-1}\, s^{-1}}$,
the bottom friction is $\mu=4\times10^{-8}\, {\rm s^{-1}}$, and the lateral eddy viscosity is $\nu=50\, {\rm m^2\, s^{-1}}$.
The asymmetric wind curl forcing, which drives the double-gyre ocean circulation, is\\
\[
\displaystyle
F_{\rm w}=\left\{ \begin{array}{ll}
\displaystyle
-1.80\, A\,\tau_o\sin\left(\pi y/y_0\right), & y\in[0,y_0), \\
\displaystyle
{\color{white}-}2.22\,A\,\tau_o\sin\left(\pi (y-y_0)/(L-y_0)\right), & y\in[y_0,L],\\
\end{array}\right.
\]
with the wind stress amplitude $\tau_0=0.03\, {\rm N\, m^{-2}}$ and the tilted zero forcing line
$y_0=0.4L+0.2x$, $x\in[0,L]$; $A=\pi/(L\rho_1H_1)$, $\rho_1=1000\,\rm kg\cdot m^{-3}$ is the top layer density,
and $H_1=250\, \rm m$ is the top layer depth.
The computational domain $\Omega=[0,L]\times[0,L]\times[0,H]$ is a closed, flat-bottom basin with $L=3840\, \rm km$, and the total depth
$H=H_1+H_2+H_3$ given by the isopycnal fluid layers of depths (top to bottom): 
$H_1=0.25\, \rm km$, $H_2=0.75\, \rm km$, $H_3=3.0\, \rm km$. 

The PV anomaly $\boldsymbol{q}$ and the velocity streamfunction $\boldsymbol{\psi}$ are coupled through the system of elliptic equations:\\
\begin{equation}
\boldsymbol{q}=\nabla^2\boldsymbol{\psi}-{\bf S}\boldsymbol{\psi} \, ,
\label{eq:pv}
\end{equation}
with the stratification matrix\\
\[
{\bf S}=\left(\begin{array}{lll}
   {\color{white}-}1.19\cdot10^{-3} & -1.19\cdot10^{-3}                  &  {\color{white}-}0.0                  \\
  -3.95\cdot10^{-4}                 &  {\color{white}-}1.14\cdot10^{-3}  & -7.47\cdot10^{-4}                  \\
          {\color{white}-}0.0          & -1.87\cdot10^{-4}                  &  {\color{white}-}1.87\cdot 10^{-4} \\
\end{array}\right).
\]
\noindent
The stratification parameters are given in units of $\rm km^{-2}$ and chosen so that the first and second Rossby deformation 
radii are $Rd_1=40\, {\rm km}$ and $Rd_2=23\, {\rm km}$, respectively.
These parameters are typical for the North Atlantic,
as they allows to simulate a more realistic 
but yet idealized Gulf Stream, compared to other QG configurations.

System~(\ref{eq:pve})-(\ref{eq:pv}) is augmented with the mass conservation constraint~\cite{McWilliams1977}:\\
\begin{equation}
\partial_t\iint\nolimits_{\Omega}(\psi_j-\psi_{j+1})\ dydx=0,\quad j=1,2
\label{eq:masscon}
\end{equation}
with a zero initial condition and a partial-slip boundary condition~\cite{HMG_1992}:\\
\begin{equation}
\left(\partial_{\bf nn}\boldsymbol{\psi}-\alpha^{-1}\partial_{\bf n}\boldsymbol{\psi}\right)\Big|_{\partial\Omega}=0 \, ,
\label{eq:bc}
\end{equation}
where $\alpha=120\, {\rm km}$ is the partial-slip parameter, and $\bf n$ is the normal-to-wall unit vector.

Before proceeding to the hybrid QG model, we describe how the reference data $\{\mathbf{q}_i,\mathbf{v}_i\}_{i\in[1,N]}$
is computed. We first spin up the QG model for a period of 50 years on a high-resolution 
grid $513\times513$ (grid step is $dx=dy=7.5\, {\rm km}$) until the solution is 
statistically-equilibrated. Then,
the model runs for another 2 years, and its 2-year solution is 
projected point-to-point onto a coarse grid $129\times129$ 
(grid step is $dx=dy=30\, {\rm km}$). 
The first of these 2 years is used as the reference solution, while the second is retained 
for hybrid model validation.
The QG solution computed at the resolution $dx=dy=30\, {\rm km}$ 
we call the modelled solution; it is used for comparison with the HP solution.

The influence of resolution on both comprehensive and idealized ocean models is well 
established: it significantly alters flow dynamics in low-resolution simulations compared 
with high-resolution, resulting in the loss of both large- and small-scale structures that are 
nevertheless resolved at low resolution.
One reason the low-resolution solution lacks large- and small-scale features of the reference flow
is that the model dissipates too much energy, thereby suppressing vortogenesis and 
the inverse energy cascade -- processes essential for the development of coherent 
large-scale flow structures. The energy-aware hybrid model, discussed in the next section,
controls the low-resolution model energy on specific scales, thereby
improving the energy of the modelled low-resolution solution towards that of the reference.
This, in turn, restores both vortogenesis and the development of large-scale flow structures.

\section{The hybrid quasi-geostrophic model}
To develop the hybrid QG model to be used together with the DA methodology, we 
add the stochastic velocity corrector~\eqref{eq:dw} and the compensating forcing 
$\mathbf{G}$ to the QG model~\eqref{eq:pve}:\\
\begin{equation}
d q^h_j+(\mathbf{u}_j dt+\mathbf{A}(\mathbf{u}_j,\mathbf{v}_j))\cdot\nabla q^h_j=(F+\mathbf{G}(q^h_j,q_j))dt,
\quad j=1,2,3\,,
\label{eq:pve_hybrid}
\end{equation}
where $\mathbf{G}(q^h_j,q_j):=\eta(\mathbf{M}(q^h_j,q_j)-q^h_j)$, and the superscript $h$ denotes the hybrid solution. 
The system of elliptic equations~\eqref{eq:pv}, 
the mass conservation constraint~\eqref{eq:masscon}, and the boundary condition~\eqref{eq:bc} 
remain unchanged.

For the purpose of this study, we use the same setup for the hybrid model as in~\cite{SC2024_J1} and
decompose the reference data into two sets (not shown): 
the first one comprises all spatial
scales within the range $s_1\in[30,300)\, {\rm km}$, while the second includes scales 
in the range $s_2\in[300,3840]\, {\rm km}$. This division is justified by the fact that
the coarse grid adequately resolves scales larger than $300\, {\rm km}$.
Scales smaller than $300\, {\rm km}$ are considered unresolved 
according to the scale resolution criterion, which requires at least 10 grid points per 
wavelength.
We note that this two-scale decomposition may not be sufficient for other GFD models, 
as accurate resolution of specific scales does not necessarily ensure proper functioning of 
intra- and inter-scale energy transfers. 
In such cases, a finer scale decomposition may be necessary.

The next step is to calculate the reference energy band for the potential energy\\
\begin{equation}
P:=\frac12\sum\limits^2_{i=1}\frac{H_{i} 
S_{i2}}{L^2 H}\iint\limits_{\Omega}(\psi_{i+1}-\psi_{i})^2\, dy dx \,, \quad H=\sum\limits^3_{i=1}H_{i} \,,
\label{eq:P}
\end{equation}
and kinetic energy 

\begin{equation}
K:=\frac12\sum\limits^3_{i=1}\frac{H_{i}}{A}\iint\limits_{\Omega}(\nabla\psi_{i})^2\, dy dx \,.
\label{eq:K}
\end{equation}
The lower and upper boundary of this energies (calculated 
from the 1-year reference record) are
$K(\mathbf{q})\in[76,90]$ and $P(\mathbf{q})\in[487,499]$, respectively. 
These boundaries are used in the optimization method to search for the scale amplitudes 
$\{\lambda_s, \gamma_s\}$, $s\in[1,2]$.

\begin{figure}[h]
\includegraphics[scale=0.6]{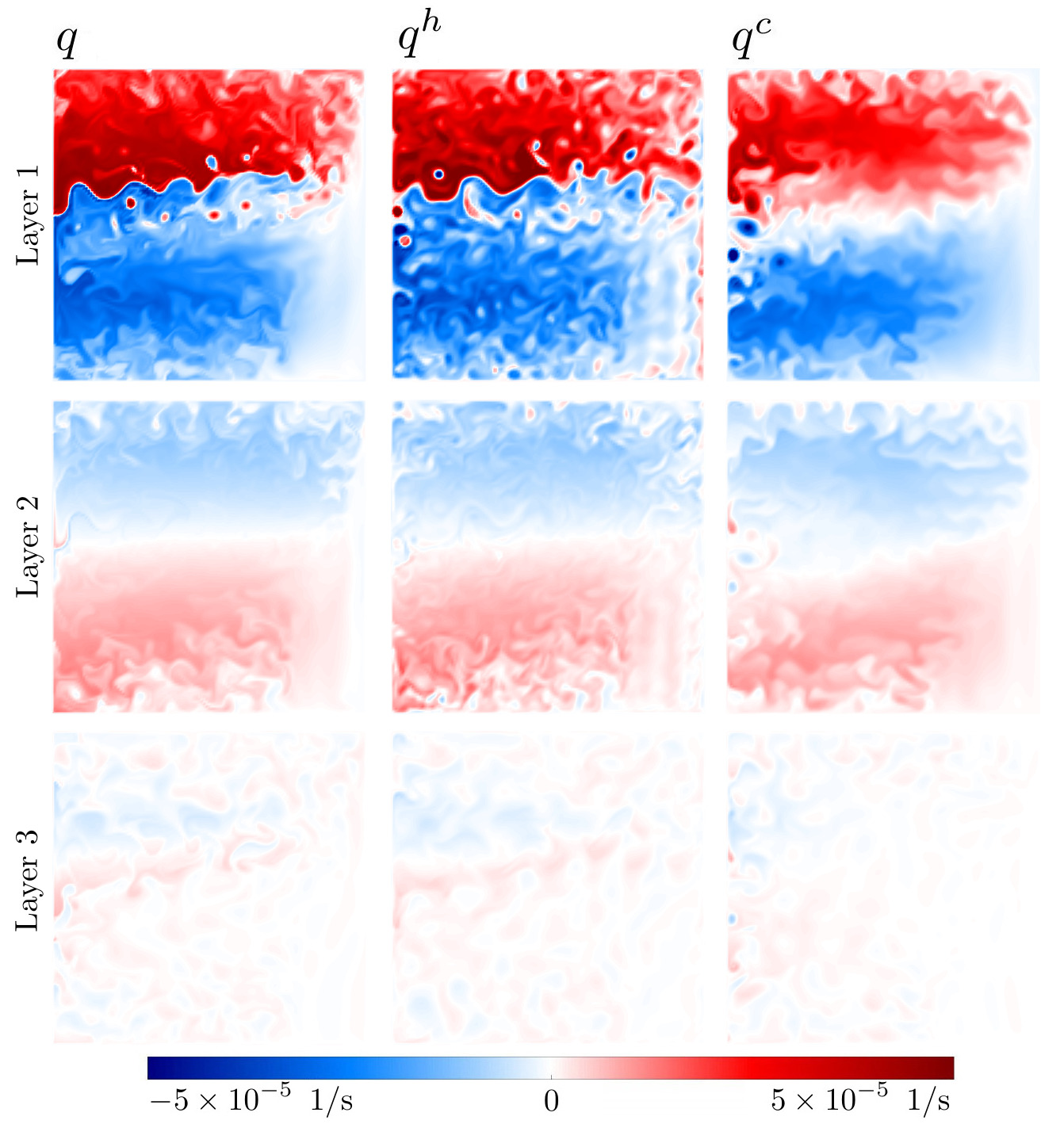}
\caption{Shown are typical snapshots of PV
for the reference $q$ (left), hybrid $q^h$ (middle), and modelled $q^c$ (right) solutions
in three layers;
all solutions are presented at the resolution $dx=dy=30\, {\rm km}$. The impact of resolution is clear:
the modelled solution has neither the large-scale jet nor small-scale vortices (which are, however, resolved
at this resolution), while the hybrid model reproduces both the jet and vortices.}
\label{fig:q1_q1c}
\end{figure}

We compare the ability of the coarse-grid QG model and its hybrid version
to reproduce high-resolution reference flow features. 
Given the reference data, its multi-scale decomposition, and the reference energy band, we 
run the hybrid QG model~\eqref{eq:pve_hybrid} over a two year period, using only the first year of the reference data. 
For the purpose of this study, it suffices to focus on the surface layer dynamics, 
as it is significantly more energetic than the lower layers and rich in large- and small-scale flow features
(figure~\ref{fig:q1_q1c}).

As shown in Figure~\ref{fig:q1_q1c}, the hybrid model 
reproduces reference flow features (the jet and the vortices)
and also keeps the total energy of the hybrid solution within the reference energy band (not shown),
whereas the QG model fails to maintain these features due to rapid energy dissipation.
It is achieved by controlling the energy at specified 
spatial scales in the hybrid model that keeps the hybrid solution in the neighborhood
of the reference phase space. 

In this work we take one step further and apply 
the ensemble-based DA methodology proposed in~\cite{CCHWS2019_1,CCHPS2020_J2} to make the hybrid solution 
even closer to the reference.
To systematically assess the benefits of this approach, we first apply the DA methodology 
to the standard (non-hybrid) QG model. This serves as a baseline for evaluating 
how well data assimilation alone can recover reference flow characteristics, 
and highlights the limitations of using DA without energy-aware hybridization. 
By comparing results from the standard QG model to those from the hybrid QG model,
each equipped with the same DA framework, we can clearly identify the added value 
provided by the hybrid strategy.

\section{Data assimilation for the quasi-geostrophic model \label{sec:qg_da}}
We first evaluate the effect of ensemble-based DA on the
QG model~\eqref{eq:pve} in the absence of hybridization. The DA implementation 
follows~\cite{CCHWS2020_4,CCHWS2019_3} with the velocity corrector~\eqref{eq:dw2}, 
thus leading to the SALT formulation~\cite{holm2015variational}:
\begin{equation}
    d q^c_j+(\mathbf{v}_j dt+\widetilde{\mathbf{A}})\cdot\nabla q^c_j=F dt\quad j=1,2,3\,.
\label{eq:pve_salt}
\end{equation}

\begin{figure}[h]
\hspace*{-1.25cm}
\includegraphics[scale=0.175]{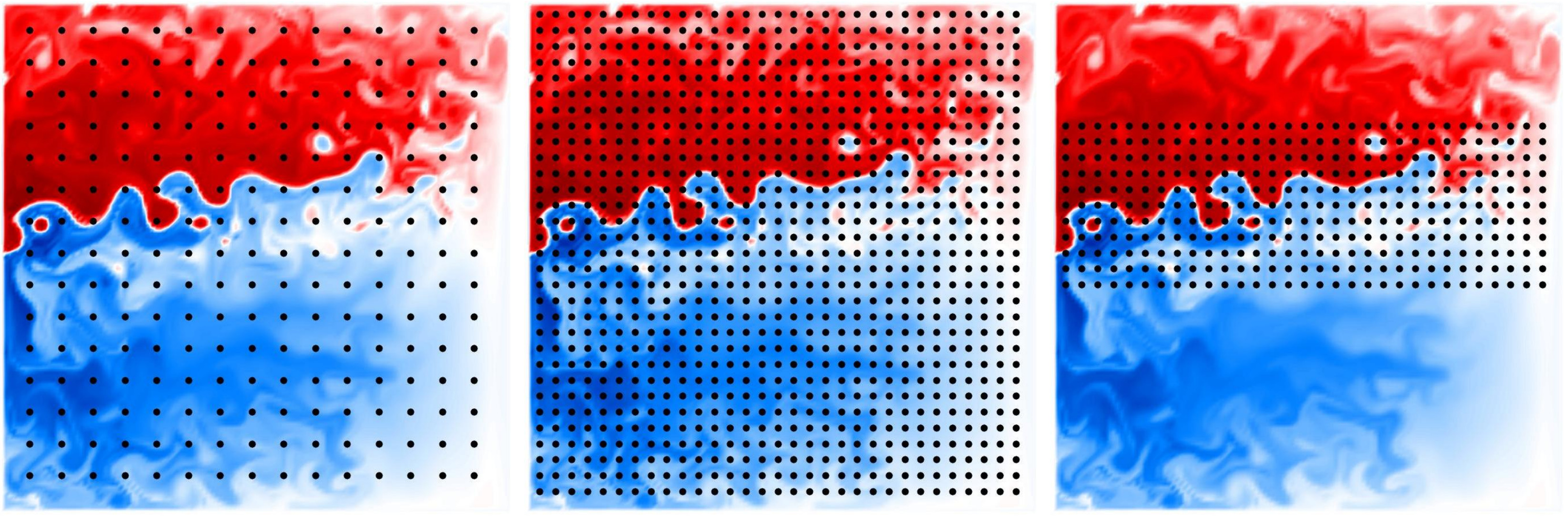}
\caption{Shown are locations of weather stations (black dots) on the surface for different 
data grids (left to right): $15 \times 15,\ 31 \times 31,\ 11 \times 31$.
The weather station locations are the same in the second and third layers (not shown).}
\label{fig:ws}
\end{figure}

To assess the impact of DA methodology and observational network design on the solution, 
we conduct a series of ensemble experiments with varying data grids. 
Figure~\ref{fig:ws} illustrates the locations of weather stations for the different assimilation 
grids: $G_d = \{3\times 15 \times 15,\ 3\times 31 \times 31\}$ span 
the entire computational domain and $G^*_d = \{3\times 11 \times 31\}$
is Gulf-Stream-focused, i.e. it samples only the most energetic region of the flow dynamics.
We hypothesize that assimilating data from the most energetic region, which
is not even present in the low-resolution model, 
should enable the hybrid solution to retain high accuracy, even though only a small fraction of 
the domain is observed. 
In all DA experiments, the initial conditions for the ensemble run of the QG model are 
generated by perturbing the reference solution 
at the initial time, ensuring that all simulations start from a small neighborhood 
of the reference state.
Also note that we deliberately skip Algorithms 1 and 2, as Algorithm 3 
(denoted as DA-A3 in the figures)
is the most accurate among those presented in this study~\cite{CCHWS2020_4,CCHWS2019_3}.
%

The performance of DA is measured using the tracking error, 
ensemble bias, and the ensemble spread. These metrics capture, respectively, 
the average deviation from the reference, the systematic offset of the ensemble mean, 
and the dispersion among ensemble members.
Given an ensemble of solutions $\widehat{\mathbf{q}}^{(i)}$, $i = 1, \ldots, N$,
the tracking error is defined as the mean relative $l_2$-norm error over the ensemble:\\
\begin{equation}
\langle{\rm RelErr}(\widehat{\mathbf{q}})\rangle:=\frac1N\sum_{i=1}^N
\frac{\|\widehat{\mathbf{q}}^{(i)}-\mathbf{q}\|_2}{\|\mathbf{q}\|_2},
\label{eq:re}
\end{equation}
where $\mathbf{q}$ is the reference solution.

The ensemble bias is defined as\\
\begin{equation}
{\rm Bias}(\widehat{\mathbf{q}}):=\langle\widehat{\mathbf{q}}\rangle-\mathbf{q}\,,
\label{eq:bias}
\end{equation}
where $\langle\cdot\rangle$ denotes the ensemble mean. 

The ensemble spread is defined as the ensemble standard deviation:\\
\begin{equation}
{\rm Spread}(\widehat{\mathbf{q}}):=\sqrt{\frac1N\sum\limits_{i=1}^N\left(\widehat{\mathbf{q}}^{(i)}-\langle\widehat{\mathbf{q}}\rangle\right)^2}\,,
\label{eq:spread}
\end{equation}
where the superscript $i$ refers to the $i$-th ensemble member.

\begin{figure}[h]
\centering
\includegraphics[scale=0.3]{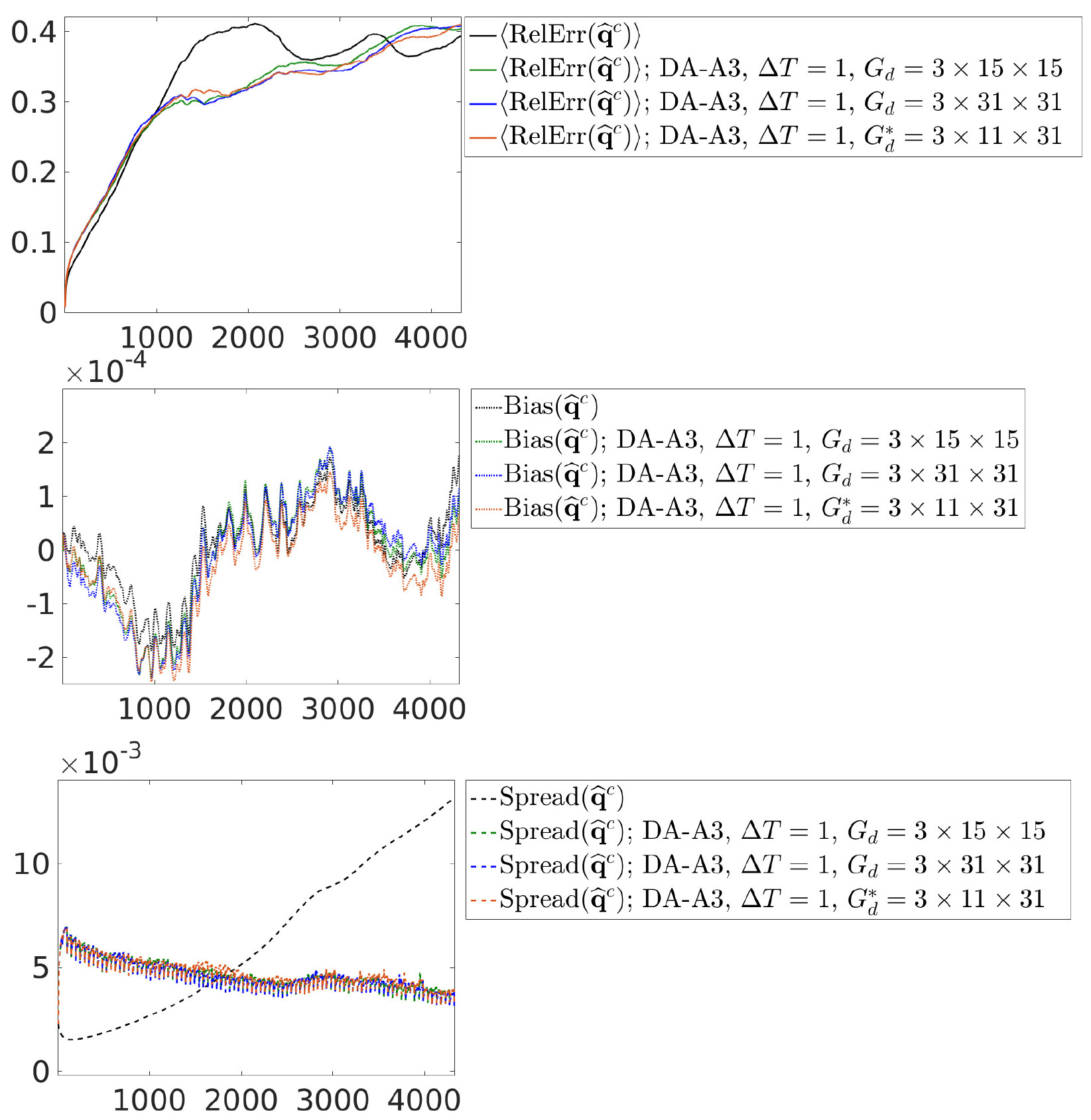}
\caption{Shown is the evolution of the tracking error (top), 
bias (middle), and spread (bottom) for 
the modelled solution $\mathbf{q}^c$ without DA and with DA (using Algorithm 3, denoted as DA-A3)
on different data grids, $G_d$, and for the DA step $\Delta T=1$ day.
}
\label{fig:re_bias_spread_dt1__no_hybrid}
\end{figure}

As seen in Figure~\ref{fig:re_bias_spread_dt1__no_hybrid},
the QG model with DA quickly hits an error qualitatively 
similar to that of the QG model without DA.
While DA successfully suppresses ensemble divergence and constrains bias, it cannot compensate 
for model errors that systematically drive the forecast away from the reference phase space. 
As the coarse-grid QG model lacks key physical processes and fails to reproduce the correct 
phase-space dynamics, DA corrections become transient: any improvement achieved at assimilation 
times is rapidly lost between updates as the model's inherent errors reassert themselves.
The results are similar for full-domain and Gulf-Stream-focused grids, 
thus suggesting that network coverage alone cannot compensate for the model error 
in this configuration.

\newpage
\begin{figure}[h]
\centering
\includegraphics[scale=0.28]{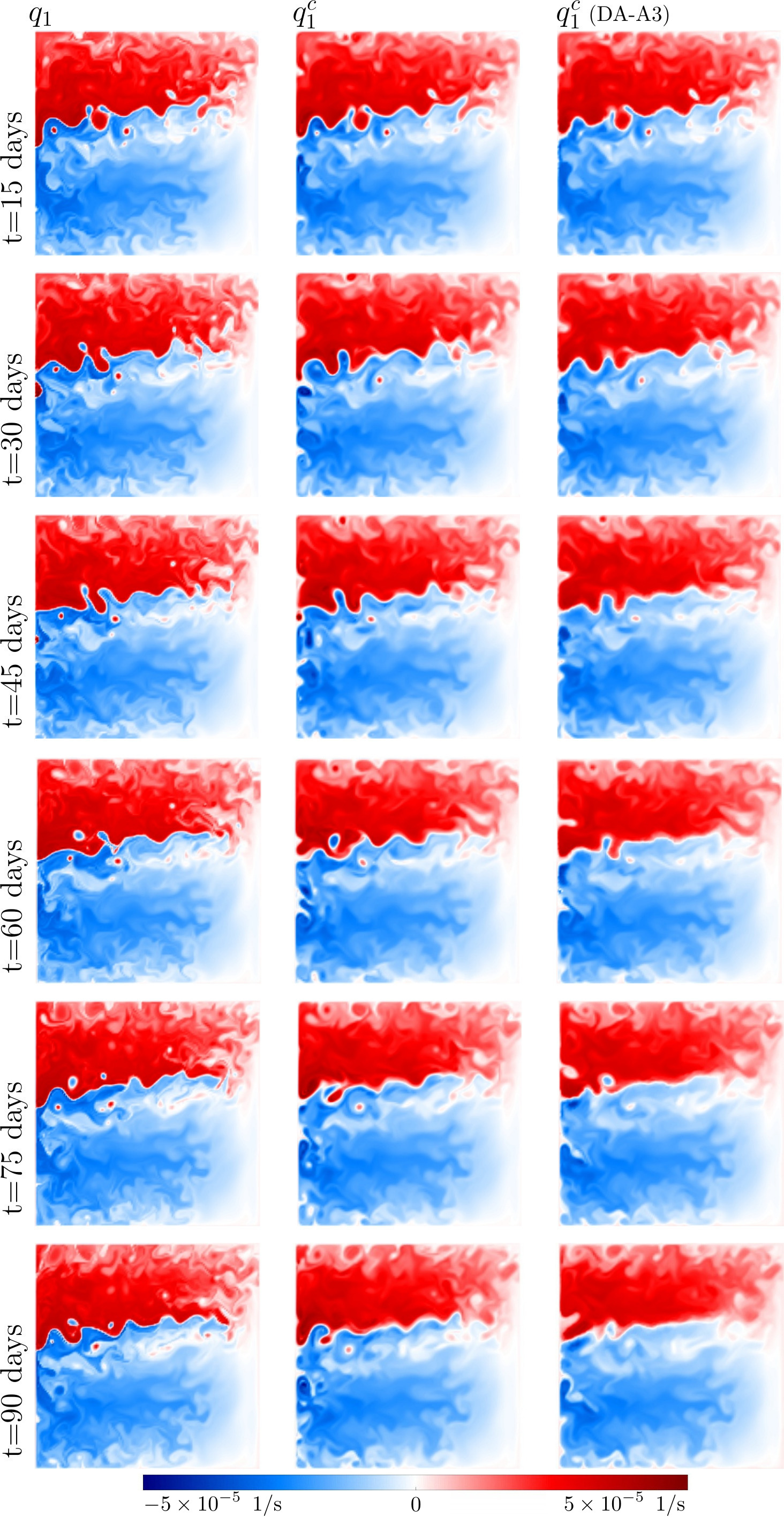}

\caption{Shown is the evolution of the reference solution $q_1$ (left column), 
a randomly chosen ensemble member of the modelled solution $q^c_1$ without DA (middle column),
and a randomly chosen ensemble member of the modelled solution  $q^c_1$ with DA (right column).
Observations are assimilated from the grid $3\times31\times31$ every $\Delta T=1$ day.
}
\label{fig:ref_modelled_modelledDAws33x33}
\end{figure}

\clearpage
The time-evolution comparison reveals that, for the coarse-grid QG model, daily assimilation
from the grid $3\times31\times31$ does 
not improve, and in some instances degrades, the dynamical realism of ensemble members (Figure~\ref{fig:ref_modelled_modelledDAws33x33}). 
Relative to the free ensemble run (no DA), the DA run shows an over-smoothed 
frontal structure, significantly reduced mesoscale variability, and eventual loss of coherent vortices, 
even at early times. 
This suggests a critical limitation of assimilation in the presence of severe 
structural model error. Because the coarse model's phase space 
does not encompass the dynamics of the reference system, assimilation increments are 
dynamically inconsistent and rapidly dissipated or distorted by the model's own biases.
Rather than correcting errors, DA accelerates the collapse toward a biased, low-energy state. 
These results underline that, with a structurally flawed model, DA can be counterproductive, 
and that meaningful improvement requires a more accurate model.
By structurally flawed model, we mean a model whose flow dynamics, 
numerical resolution, or parameterizations are so different from the reference system that:
the model's phase space is far away from the reference phase space;
key physical processes are missing, misrepresented, or dissipated;
model errors are systematic, not random.

\noindent
{\bf Limitations of purely stochastic correction}.
While the EOF-based corrector, $\widetilde{\mathbf{A}}$, injects spatial variability designed 
to mimic the effect of unresolved dynamics onto the resolved, it does not distinguish between the physical sources of model 
error (e.g. energy imbalance,  incorrect representation of large-scale circulation, etc.) and 
random, small-scale variability (e.g. fluctuations due to unresolved turbulent eddies or 
stochastic forcing).
This can result in suboptimal or even counterproductive updates: the ensemble may spread, 
but its mean does not systematically approach the reference state. In the absence of a targeted 
deterministic correction, the ensemble can diverge, and flow features such as fronts, jets, and 
eddies are poorly represented thus impairing the backward energy cascade, which in turn 
leads to the decay of large-scale structures.

\noindent
{\bf Missing energy-aware control and error correction}.
The original DA scheme (with $\mathbf{G}=0$) lacks any deterministic, dynamically-adaptive 
correction to the model's large-scale, low-frequency error. In the hybrid framework, 
the $\mathbf{G}$-term is critical: it nudges the model state toward the reference phase space 
by dynamically correcting the energy content at targeted scales, 
thereby compensating for systematic biases and drift inherent to the low-resolution model.
When $\mathbf{G}=0$, these biases are not controlled, and stochastic perturbations alone 
(even if based on calibrated EOFs) are happened to be insufficient to systematically reduce the tracking error 
over time. Instead, the QG model quickly drifts away from the reference phase space.

In summary, the combination of omitted energy control and reliance solely on 
stochastic, EOF-based corrections is not sufficient to ensure accurate tracking of the reference 
solution and stability 
of the DA method in the standard QG model. This is especially true in realistic, high-variance regimes, 
where systematic errors dominate and cannot be mitigated by advection velocity corrections alone.
This observation motivates our focus on hybrid DA approaches that combine energy-aware 
corrections with stochastic ensemble methods, as developed and tested in the remainder of 
this study.

\section{Data assimilation for the hybrid quasi-geostrophic model \label{sec:qg_hybrid_da}}
This section extends our analysis by applying the ensemble-based DA methodology 
to the hybrid QG model. This approach allows us to directly compare 
the performance of the hybrid and non-hybrid models under identical assimilation settings
and to quantify the added value of hybridization for accurate and physically consistent 
flow dynamics.

\begin{figure}[h]
\centering
\includegraphics[scale=0.175]{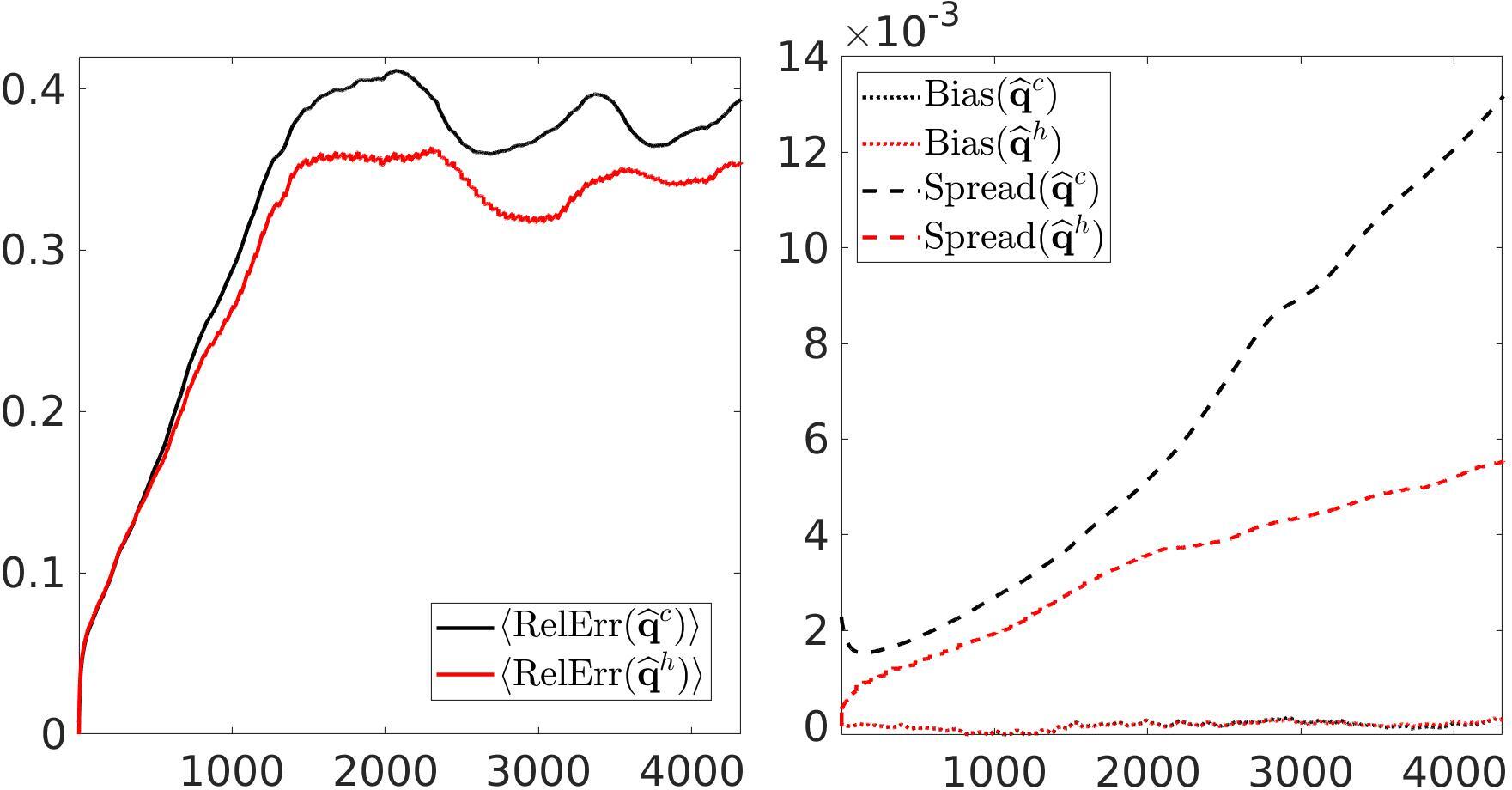}
\caption{Shown is the evolution of the tracking error (left), 
bias and spread (right) for the modelled solution$\mathbf{q}^c$ and the hybrid solution $\mathbf{q}^h$;
there is no data assimilation applied.}
\label{fig:re_bias_spread_qc_qh}
\end{figure}

First, we compare an ensemble run of the coarse-grid QG model~\eqref{eq:pve}
with an ensemble run of the hybrid model~\eqref{eq:pve_hybrid} (Figure~\ref{fig:re_bias_spread_qc_qh}).
As shown in Figure~\ref{fig:re_bias_spread_qc_qh}, the hybrid model yields more accurate predictions
and also exhibits slower ensemble divergence compared to the classical QG model. 
Both models display small ensemble bias.
It is worth noting that increasing the nudging parameter $\eta$,
can further reduce $\langle{\rm RelErr}(\widehat{\mathbf{q}}^h)\rangle$.
A systematic study of this parameter's influence, while potentially 
reinforcing the hybrid model's advantage, lies outside the scope of the present work, 
which is restricted to assessing the impact of the DA methodology within the 
framework of energy-aware hybrid modeling.

\newpage
\begin{figure}[h]
\centering
\includegraphics[scale=0.26]{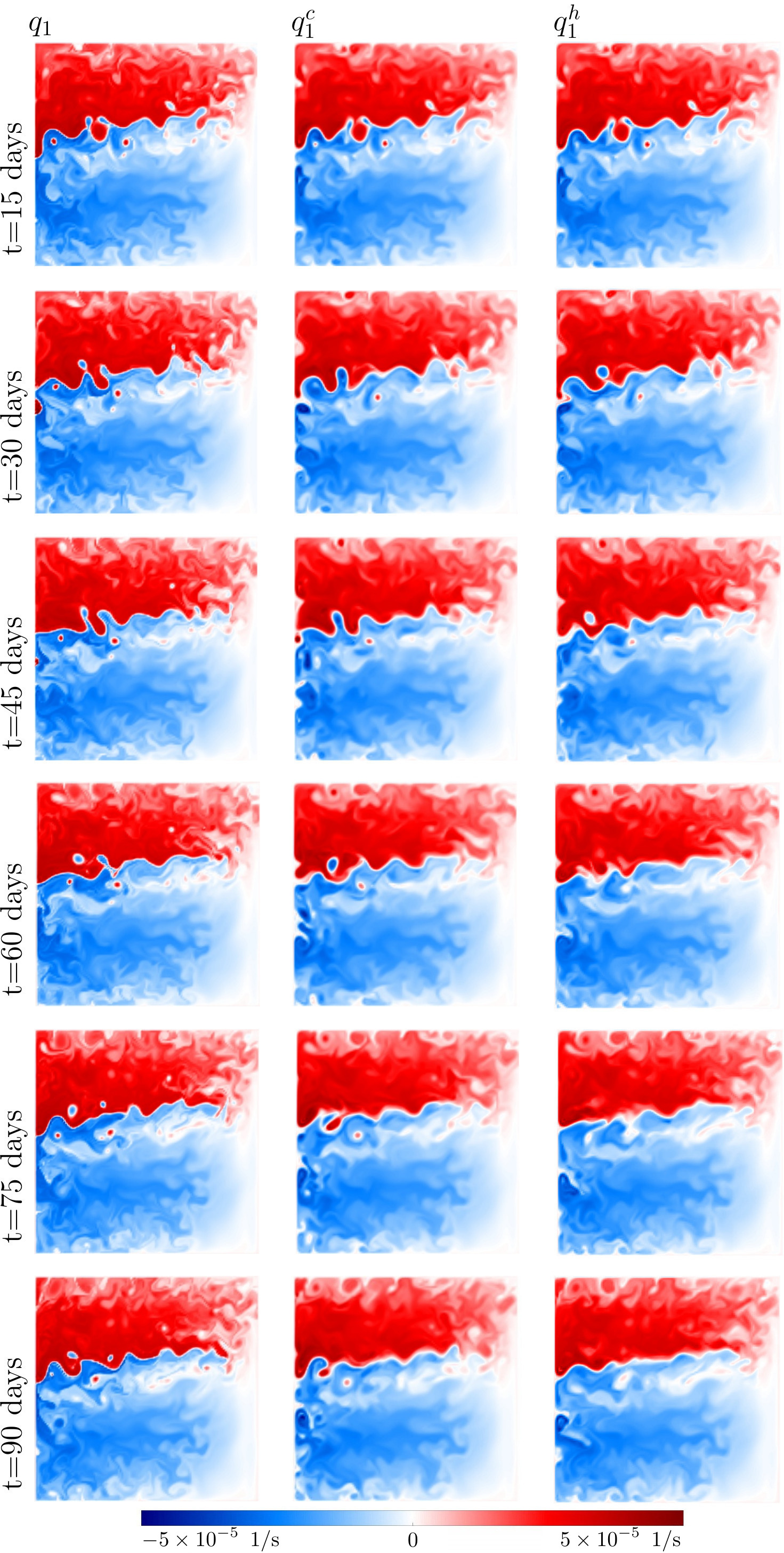}

\caption{
Shown is the evolution of the reference solution $q_1$ (left column), 
a randomly chosen ensemble member of the modelled solution $q^c_1$ (middle column),
and a randomly chosen ensemble member of the hybrid solution  $q^h_1$ (right column).
Both models exhibit similar large-scale jets over 90 days because the standard QG model, 
when started from the reference initial condition, drifts very slowly towards its 
lower-energy phase space (multi-year timescale). 
Quantitative differences, most apparent in mesoscale structures, are shown in Figure~\ref{fig:re_bias_spread_qc_qh}.
}
\label{fig:ref_modelled_hybrid}
\end{figure}

\clearpage
Although Figure~\ref{fig:ref_modelled_hybrid} shows broadly similar large-scale structures
for the QG and its hybrid version, quantitative metrics in
Figure~\ref{fig:re_bias_spread_qc_qh} make the distinction unambiguous. 
The hybrid model maintains a consistently lower relative error than the standard QG,
with much smaller ensemble spread and comparable bias throughout the simulation. 
%
%
The modest visual difference over 90 days is partly explained by the fact that 
the standard QG model, when started from the reference initial condition, drifts very 
slowly towards its lower-energy phase space, this process takes almost 5 years 
as found in~\cite{SC2024_J1}.
Over the short integration period shown, the hybrid's advantage is therefore most 
visible in objective error metrics rather than in large-scale visual patterns.

\begin{figure}[h]
\centering
\includegraphics[scale=0.3]{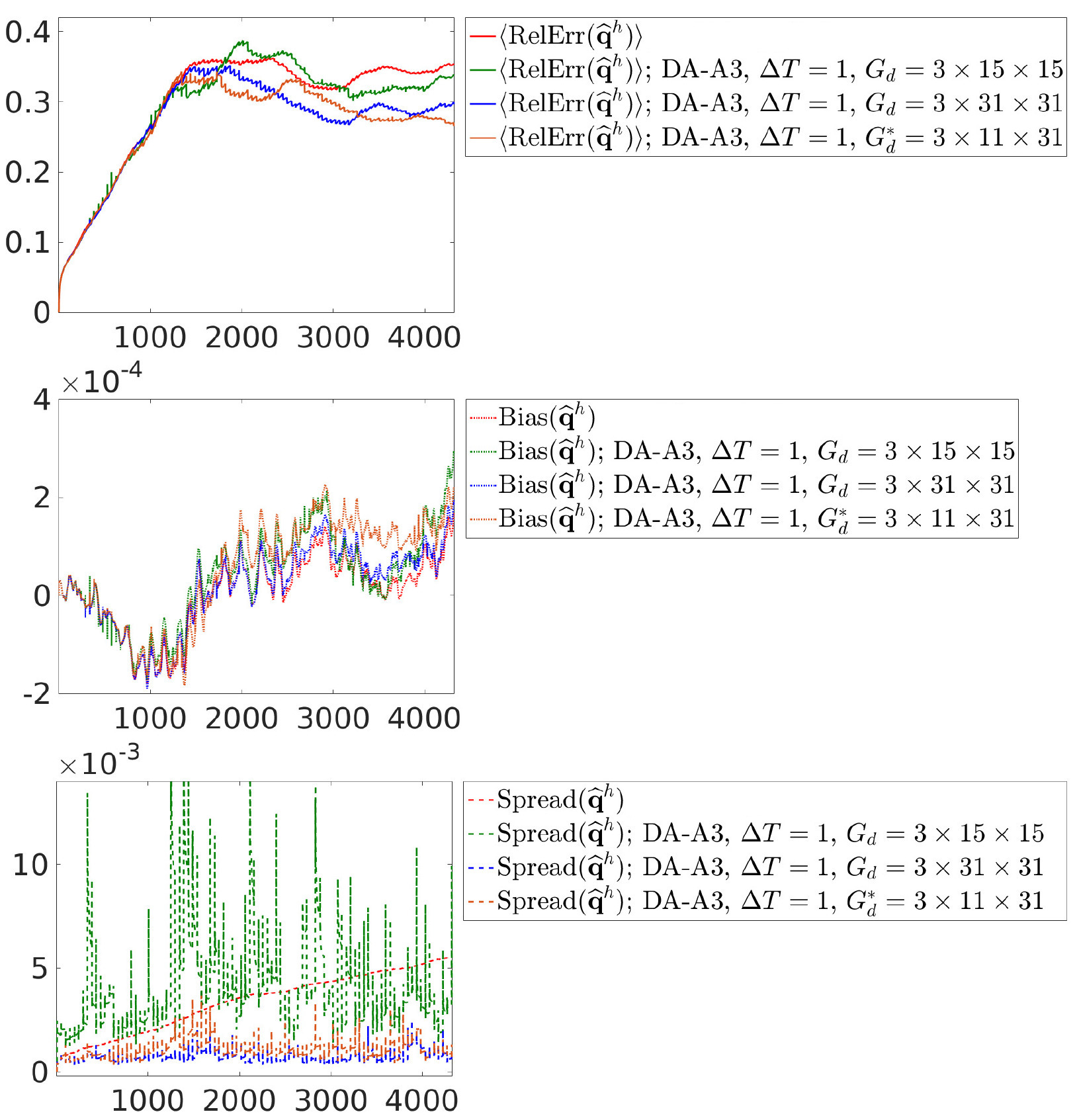}
\caption{Shown is the evolution of the tracking error (top), 
bias (middle), and spread (bottom) for 
the hybrid $\mathbf{q}^h$ solutions, using the DA
algorithm 3 (DA-A3) on different data grids, $G_d$, and for the DA step $\Delta T=1$ day.
}
\label{fig:re_bias_spread_dt1}
\end{figure}

In our next examples, we analyse how the DA methodology, based on Algorithm~\ref{alg:nudging},
affects the hybrid solution.
First, we fix the DA step $\Delta T=1$ day and study how different data grids
influence the results (Figure~\ref{fig:re_bias_spread_dt1}).
As seen in Figure~\ref{fig:re_bias_spread_dt1}, the size of data grid significantly 
affects the accuracy. For the smallest grid $G_d=3\times15\times15$ covering the whole computational
domain, the DA method shows
almost no improvement compared with the hybrid solution 
(compare red and green curves in Figure~\ref{fig:re_bias_spread_dt1}), the bias is on par 
with the that of the hybrid solution, and the spread manifests substantial fluctuations but with no
uptrend though (as for the hybrid model without DA). The use of a larger data grid, 
$G_d=3\times31\times31$, leads to a systematically more accurate solution 
(compare red and blue curves in Figure~\ref{fig:re_bias_spread_dt1}) and to much less
uncertainty within the ensemble. The most interesting finding though is the results for 
the grid focused on the Gulf Stream region $G^*_d=3\times11\times31$. Even though this grid 
does not cover the whole domain, it offers the error and the ensemble uncertainty 
which are on the same level with that of the densest grid used
(compare blue and brown curves in Figure~\ref{fig:re_bias_spread_dt1}). 
In stark contrast to the application of DA to the QG model, where the Gulf Stream focused 
assimilation failed, this result confirms
our hypothesis that assimilating data from the most energetic part of the flow dynamics does not
compromise the accuracy of the hybrid solution, despite only a small fraction of the domain is observed.
Doubling the ensemble size to $N=100$ does not alter these conclusions (not shown).

\clearpage

\begin{figure}[h]
\centering
\includegraphics[scale=0.3]{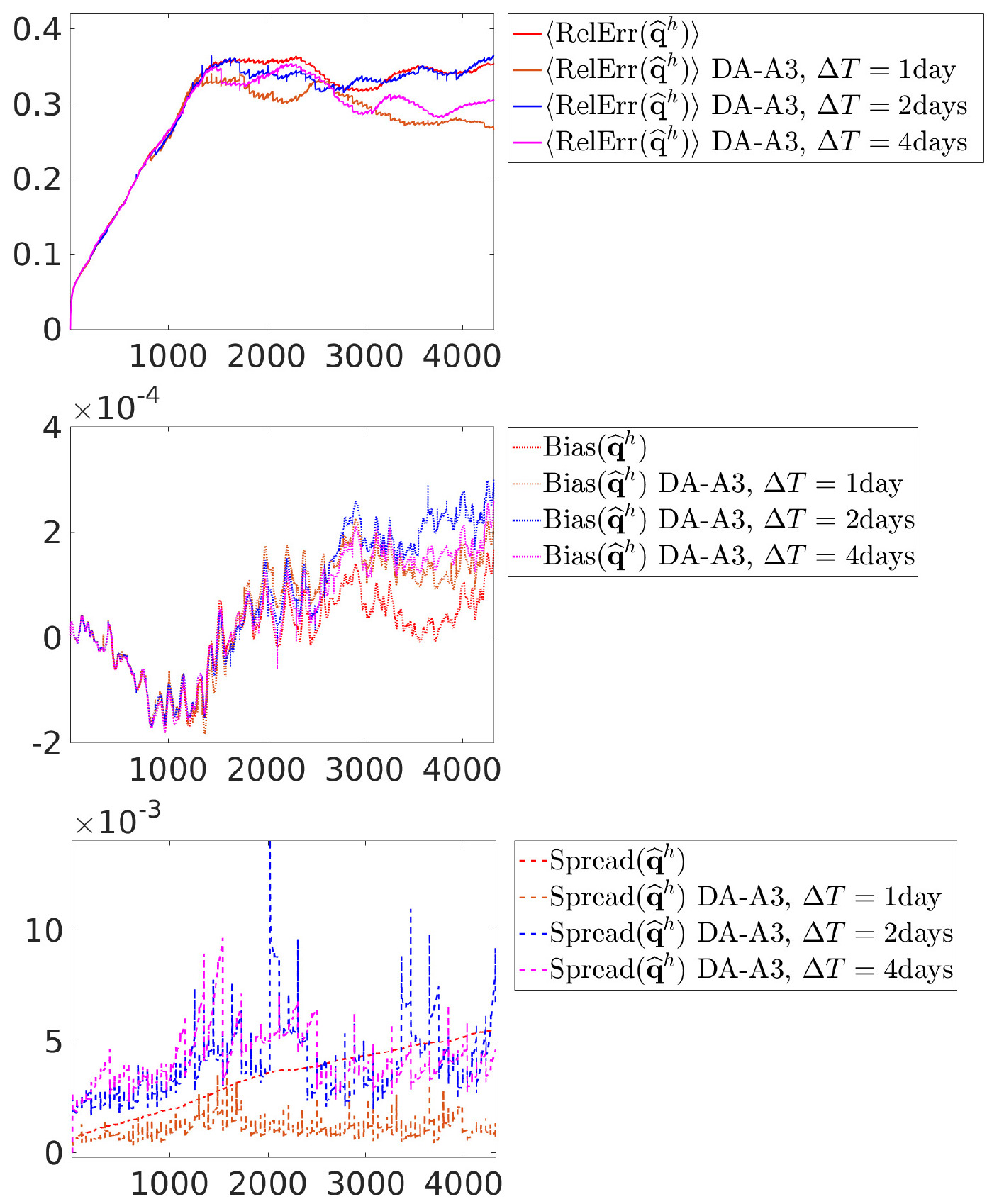}
\caption{Shown is the evolution of the tracking error (top), 
bias (middle), and spread (bottom) for the hybrid $\mathbf{q}^h$ solution, using the DA 
algorithm 3 (DA-A3) in the Gulf Stream region, $G^*_d=3\times11\times31$, 
for the DA steps $\Delta T=\{1,2,4\}$ days.}
\label{fig:re_bias_spread_dt_da}
\end{figure}

We now take the Gulf Stream focused grid, $G^*_d$, and analyse how the length of the DA 
step affects the results (Figure~\ref{fig:re_bias_spread_dt_da}). 
As shown in Figure~\ref{fig:re_bias_spread_dt_da},
shorter assimilation intervals lead to lower error. The hybrid model paired with DA outperforms the 
free running hybrid SPDE for all $\Delta T$, but the gains erode as $\Delta T$ increases, thus
showing the efficiency of the method, but only if one assimilates frequently enough.
The same conclusion is valid for the bias (as $\Delta T$ increases DA's ability to 
suppress bias decreases) and the spread, which balloons at larger $\Delta T$.

\newpage
\begin{figure}[h]
\centering
\includegraphics[scale=0.27]{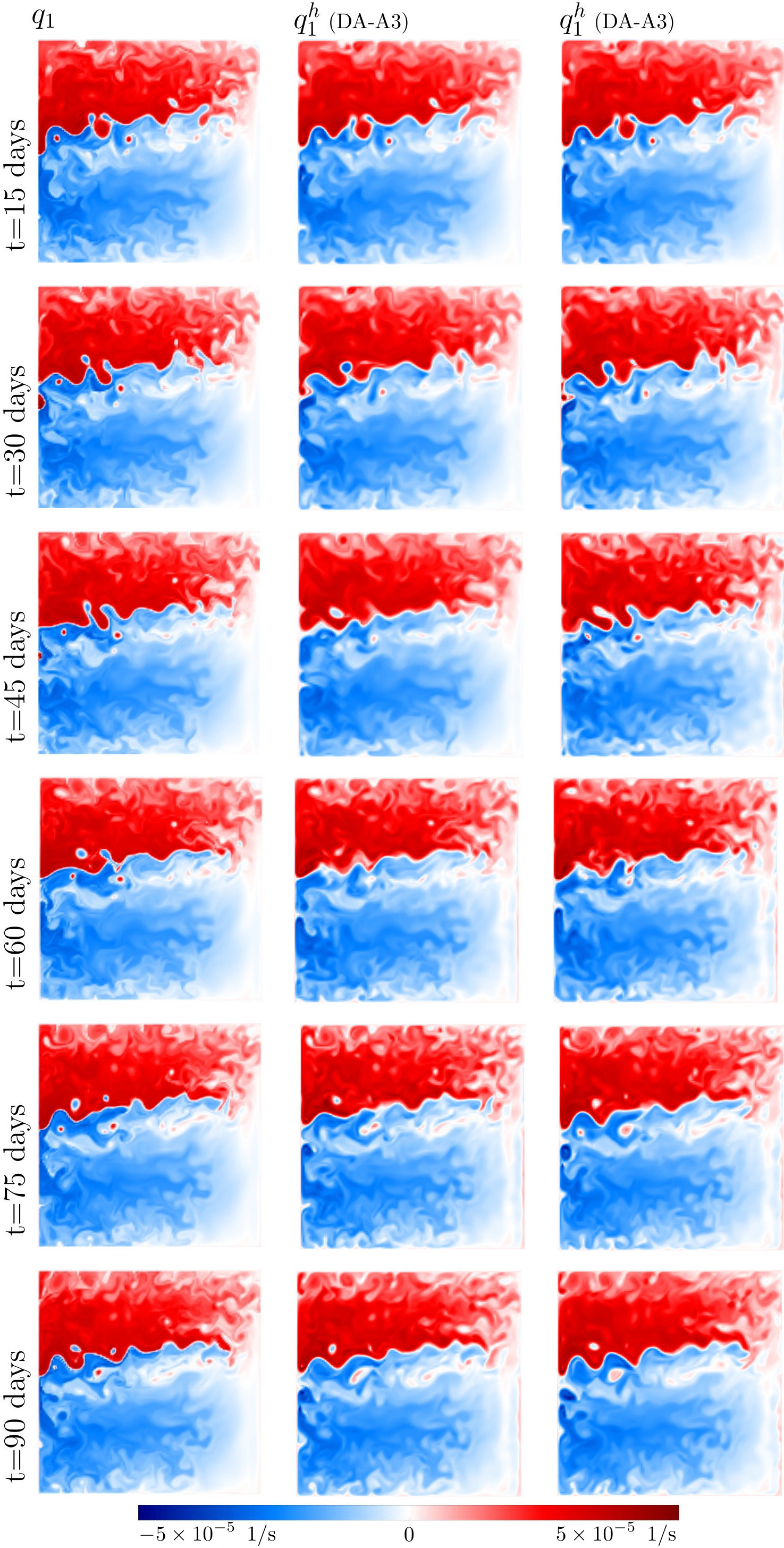}

\caption{
Shown is the evolution of the reference solution $q_1$ (left column), 
a randomly chosen ensemble member of the hybrid solution $q^h_1$, using DA-A3
on the grid $3\times31\times31$ and with $\Delta T=1$ day, (middle column),
and a randomly chosen ensemble member of the hybrid solution $q^h_1$, using DA-A3
on the Gulf Stream focused grid $3\times11\times31$ and with $\Delta T=1$ day, (right column).
}
\label{fig:ref_modelled_hybrid_DA}
\end{figure}

\clearpage

Figure~\ref{fig:ref_modelled_hybrid_DA} highlights the stable skill of the hybrid model 
when coupled with DA over 90 days. Both assimilation configurations -- 
full-domain (middle column) and Gulf Stream focused (right column) -- 
maintain close phase alignment with the reference solution (left column) and 
successfully preserve the sharpness of the Gulf Stream front, coherent vortices, and 
mesoscale filaments throughout the simulation. The similarity between the two DA 
configurations is striking: despite the reduced spatial coverage, targeted assimilation 
in the dynamically active Gulf Stream region produces results that are very close to
full-domain assimilation, particularly in maintaining the energetic frontal zone, vortogenesis,
and suppressing the large-scale drift. 
This demonstrates that, for a model whose energetics and phase-space occupancy are 
consistent with the reference system, observations placed in key 
energetic regions can achieve near-optimal tracking of the reference flow while 
substantially reducing the observational footprint. These results reinforce 
the importance of optimal observation placement and suggest that, in the hybrid modeling 
framework, targeted DA can provide a cost-effective alternative to uniformly 
distributed observations without sacrificing accuracy.

\begin{figure}[h]
\centering
\includegraphics[scale=0.3]{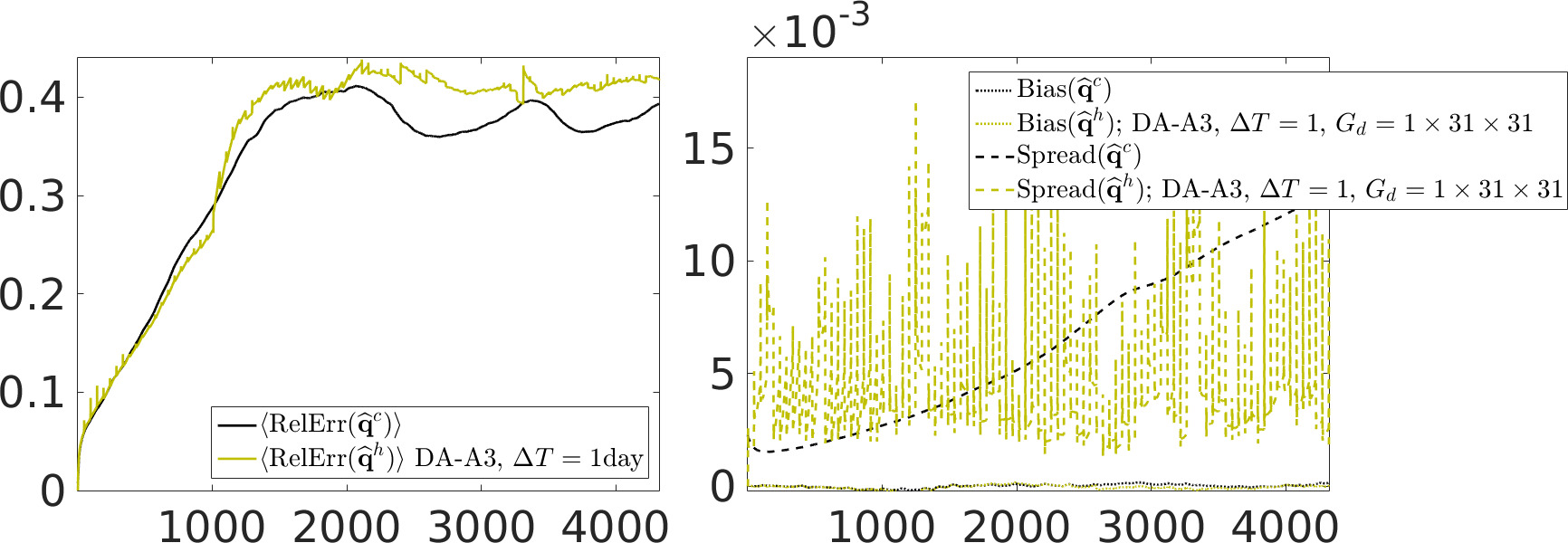}
\caption{Shown is the evolution of the tracking error (left), bias and spread (right) for the modelled solution 
$\mathbf{q}^c$ and the hybrid solution $\mathbf{q}^h$ with surface-only data assimilation 
(DA-A3, $\Delta T = 1$ day, $G_d = 1 \times 31 \times 31$). 
Despite relatively well-sampled surface grid, DA fails to improve accuracy, i.e.
the tracking error remains comparable to or slightly worse than that of the QG model without DA.
}
\label{fig:re_bias_spread_surface_only}
\end{figure}

\begin{figure}[h]
\centering
\includegraphics[scale=0.28]{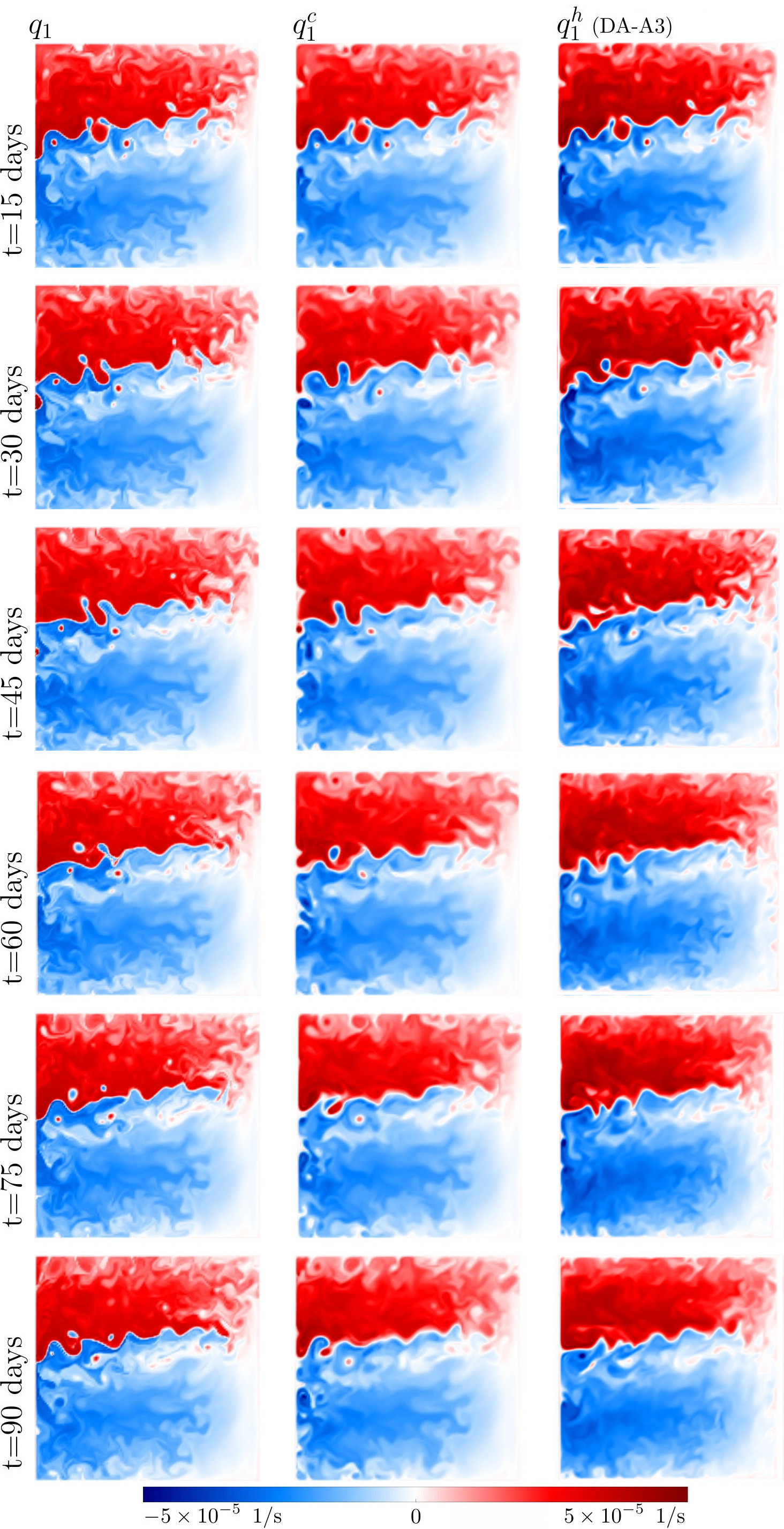}
\caption{Shown is the evolution of the reference solution $q_1$ (left column), 
a randomly chosen ensemble member of the modelled solution $q^c_1$ without DA (middle column),
and a randomly chosen ensemble member of the modelled solution  $q^c_1$ with 
surface-only data assimilation (DA-A3, $\Delta T = 1$ day, $G_d = 1 \times 31 \times 31$),
(right column).
}
\label{fig:ref_modelled_hybrid_surface_only_DA}
\end{figure}

However, when assimilation is restricted to surface observations only, the outcome is markedly different.
Surface-only DA significantly degraded the accuracy of the hybrid model,
it performs even worse than the coarse-grid model without DA (Figure~\ref{fig:re_bias_spread_surface_only}). 
Figure~\ref{fig:ref_modelled_hybrid_surface_only_DA} provides a flow-field perspective on the 
impact of surface-only DA. The reference solution (left column) maintains a sharp 
Gulf Stream front with coherent vortices and filaments throughout the 90-day period. 
The free coarse-grid model (middle column) gradually loses mesoscale variability, producing an 
overly smooth frontal zone by day 75. By contrast, the hybrid model with surface-only DA 
(right column) initially produces a sharper front than the free run, but this correction is 
not dynamically sustainable: vortices are distorted, frontal position drifts, 
and small-scale structures are degraded after 45 days. Thus, while surface increments 
temporarily improve the surface flow, they fail to enforce vertical balance, leading to 
dynamical inconsistency. This surface flow dynamics corroborates the error diagnostics in Figure~\ref{fig:re_bias_spread_surface_only},
thus confirming that surface-only DA provides no lasting improvement and can even degrade 
the hybrid solution.

\clearpage

\clearpage
This counter-intuitive result can be explained by the strong vertical coupling of the 
QG system. The surface potential vorticity is not an isolated diagnostic; 
it is dynamically linked to the subsurface layers through PV inversion and baroclinic interactions. 
When observations are assimilated in all three layers, the increments respect this coupling, 
producing dynamically consistent corrections that reinforce the hybrid model's energy-aware control.
By contrast, when assimilation is restricted to the surface layer, the imposed increments adjust 
the upper-level fields without corresponding changes in the lower layers. The result is a 
state that matches the surface data but is inconsistent with the model's vertical structure and 
energetics.

From a phase-space perspective, the surface-only increments force the hybrid model into regions of 
phase space that may look correct at the surface but do not correspond to the solutions of 
the governing equations. Because the hybrid framework is designed to maintain energy consistency 
across scales, these inconsistent corrections can generate spurious vertical shear and artificial 
imbalances in the energy distribution. The model then rejects the increments by dissipating them 
or distorting them, leading to rapid error growth. 
In effect, the assimilation becomes counterproductive: instead of nudging the system towards the 
reference phase space, it continually injects inconsistencies that accelerate divergence.

This behavior illustrates a fundamental principle of data assimilation: 
the effectiveness of observations depends not only on their number, but also on their 
dynamical consistency with the model~\cite{lorenc1986analysis,evensen2009data,reich2015probabilistic}.
Surface-only observations lead to degraded accuracy despite providing proper coverage in 
the top layer.
For strongly baroclinic flows, vertically distributed observations 
are essential. Surface-only DA fails because the hybrid model, despite its energy-aware corrections,
still requires information on subsurface dynamics to maintain phase-space alignment with the 
reference system. In this sense, the result provides a striking example of {\bf the model adequacy 
problem}: the effectiveness of DA depends not only on the assimilation algorithm but also on whether 
the observations and the model together can represent the full dynamical structure of the system.

To sum up, the numerical experiments presented in this section demonstrate that the hybrid QG 
model, when equipped with ensemble-based data assimilation, consistently outperforms 
the standard QG model across all tested scenarios. The hybrid framework not only yields 
more accurate tracking of the reference solution, but also maintains smaller ensemble divergence (compared 
to the QG model), even in the absence of data assimilation. 

A key finding is the critical role of the observation network design: 
focusing observations on the most energetic region of the flow delivers accuracy and uncertainty
reduction comparable to those achieved with full-domain grids, thus highlighting the importance
of weather station locations to maximize the impact of data assimilation, 
especially when resources for observational coverage are limited.
Furthermore, the benefits of data assimilation are most pronounced when observations are 
assimilated at shorter intervals. As the assimilation interval increases, tracking error and 
spread control deteriorate. These results underscore the importance of both observational 
design and assimilation frequency in realizing the full potential of DA in hybrid modeling.

Overall, this section provides strong evidence that combining energy-aware hybrid modeling 
with advanced ensemble-based data assimilation produces reliable results in 
complex geophysical flow regimes, even under sparse, noisy, regionally focused observations.

\section{Conclusions and discussion\label{sec:conclusions}}
In this study, we have combined both the QG model and 
the energy-aware hybrid QG model with the ensemble-based data assimilation (DA) methodology, 
which includes model reduction, tempering, jittering, and nudging, to 
improve the tracking error, bias, and spread of these models, as well as to study to what extent
the DA methodology can correct the solution.

The original DA approach, which 
relies primarily on stochastic corrections based on calibrated EOFs, was previously shown to 
perform well in the QG channel flow configuration~\cite{CCHWS2020_4,CCHWS2019_3}. 
This is why we decided to use it in this study for the standard QG model.
In the QG channel flow configuration, the flow dynamics is dominated by both strong coherent structures, 
such as jets, boundary layers, and vortices, and a range of small-scale turbulent features.
Despite this complexity, the original DA scheme
was able to achieve good tracking skill, spread, and bias control. One possible explanation is that, in this specific 
configuration, the model error structure was sufficiently captured by the leading 
EOF modes, and systematic errors (such as large-scale energy imbalance or mean flow drift) 
were either less pronounced or could be partially compensated by the stochastic perturbations.

By contrast, the Gulf Stream flow configuration, studied in this work, presents a much more 
complex and dynamically richer flow regime. Here, the model errors are dominated 
by persistent, systematic biases, such as significant energy loss at small scales, and absence of
both large- and small-scale structures (at low resolutions).
We have found that stochastic corrections alone are unable to compensate for these long-lived and 
physically structured errors (energy imbalance, mean flow error, bias in coherent structures, etc.); 
instead, the ensemble may simply diverge, or fail to evolve
within the reference phase space. In such cases, additional physically informed, energy-aware corrections 
are essential for successful data assimilation and accurate tracking of the reference flow.

In summary, the success of the original DA methodology for the QG channel flow 
and its failure in the Gulf Stream setup highlights a critical point: 
the effectiveness of the DA methodology is highly 
dependent on how well the model can represent reference flow features.

In order to address this point and apply DA to the Gulf Stream flow, our approach leverages recent progress in hybrid modeling, which imposes explicit control 
over the energy at selected scales to maintain the hybrid solution within the reference energy band.
By integrating the ensemble-based DA within this hybrid paradigm, 
we have addressed key challenges associated with low-resolution modeling, including 
the loss of critical flow features (the jet and the vortices) and the 
efficient assimilation of sparse and regionally focused observational data.
Through a set of numerical experiments, we have demonstrated several important outcomes. 

{\bf Hybrid model alone restores missing dynamics}.
The coarse-grid energy-aware hybrid model alone reproduces both the large-scale 
jet and small-scale vortices (which are present in the high-resolution reference 
solution) -- features that are notably absent in traditional QG simulations 
at the same resolution. This improvement is achieved by constraining the model's
energy within the reference energy band, ensuring the hybrid solution 
remains dynamically consistent and physically realistic.

{\bf Hybrid+DA further increase accuracy}.
We have shown that the application of the ensemble-based DA methodology
to the hybrid model further increases the accuracy of hybrid simulations. 
Systematic comparison of tracking error, bias, and ensemble 
spread indicates that the hybrid model, when coupled with DA, 
demonstrates lower tracking error and ensemble divergence relative to the standard (non-hybrid)
QG model.

{\bf Targeted observations are highly effective}.
Substantial benefits are observed when assimilating observations from the most 
energetic region of the flow -- the Gulf Stream region. Remarkably, our results reveal that targeted 
assimilation in this region delivers predictive skill and uncertainty reduction 
comparable to those obtained using full-domain observational networks, 
despite the much smaller spatial coverage. This finding highlights the efficiency and 
potential cost-effectiveness of strategically designed observation campaigns, especially in 
resource-limited settings.

{\bf Assimilation frequency matters}.
Our analysis of assimilation frequency shows that more frequent updates 
(shorter DA intervals) are essential to sustain low error and ensemble coherence. 
Infrequent assimilation leads to increased tracking error and divergence 
within the ensemble, thus underscoring the importance of both observational density and 
temporal coverage in ensemble-based DA for the hybrid model.

{\bf Surface-only assimilation is counterproductive}. 
Another notable outcome is the failure of surface-only assimilation. 
Despite providing proper coverage of the most energetic layer, such updates degraded 
the hybrid solution, which performs even worse than the coarse-grid model without DA. 
This occurs because surface increments are dynamically inconsistent with the unobserved 
subsurface layers, disrupting vertical energy balance and driving the system away from 
the reference phase space. 
In baroclinic regimes, where surface and interior dynamics are tightly coupled, 
vertically distributed observations are essential. 
This finding highlights that observation design must account for both horizontal targeting 
and vertical coverage: without the latter, assimilation can become destabilizing 
rather than corrective.
These results highlight important guidance for observation network design. Although surface 
data (e.g. satellite altimetry, drifters, etc.) are abundant and relatively inexpensive, 
our experiments show that surface-only assimilation can be counterproductive. 
Accurate state estimation requires vertically distributed information, such as from 
Argo floats, gliders, or other subsurface observing platforms, to constrain the coupled dynamics 
across layers. In practical terms, even sparse but strategically placed subsurface measurements 
can deliver disproportionate benefits, by ensuring that assimilation increments respect the 
model's vertical structure and energetics. This underscores the importance of integrating 
complementary observation types (surface and subsurface) when designing efficient observing 
systems for hybrid DA frameworks.

{\bf Model fidelity is fundamental}.
When DA is coupled with the hybrid model designed to remain within 
the reference phase space, the improvements are both significant and long lasting.
Our findings highlight a fundamental and often misunderstood limitation: 
DA is only as effective as the model it corrects. If the model cannot evolve within the reference
phase space, 
no assimilation scheme, however sophisticated, can close the gap. 
These results underscore the necessity of improving forecast model fidelity in parallel with 
advancing DA methodologies.

The contrasting performance of the QG and hybrid QG models under identical 
DA settings illustrates a broader issue in data assimilation -- the need for the 
model's phase space to be compatible with the reference flow dynamics. We refer to this 
as the \textit{Model Adequacy Problem}.
Although the term ``Model Adequacy Problem'' is not universally adopted, 
the concept is recognized in DA theory, e.g. where the forecast model is structurally 
insufficient to represent the reference dynamics, causing assimilation increments to vanish or 
mislead~\cite{Bonavita2024}.

\noindent\fbox{%
\begin{minipage}{0.92\textwidth}
       \textbf{The Model Adequacy Problem} \\
In data assimilation, the \emph{model adequacy problem} arises when the forecast model's 
set of dynamically accessible states (its phase space) does not sufficiently overlap with 
the reference phase space. In such cases, even perfect observations and an optimal assimilation 
scheme cannot sustain accurate state estimates: assimilation increments are dynamically 
inconsistent and are rapidly dissipated or distorted by the model's own biases.\\

This problem is especially severe when model errors are \textbf{structural} -- stemming from missing 
physics, coarse resolution, or misrepresented processes -- rather than \textbf{random} noise.
Resolving the model adequacy problem requires improving the forecast model itself (e.g., through hybridization, 
better parameterizations, or increased resolution) so that data assimilation can produce persistent 
corrections.

In the present study, the QG model (with DA) suffers from model 
inadequacy, whereas the energy-aware hybrid QG model alleviates it, enabling DA to deliver 
stable and accurate results.
\end{minipage}
}

This finding also explains why targeted Gulf Stream observations, ineffective in the inadequate 
QG model, achieved full-domain skill in the hybrid model once phase-space compatibility 
was restored, assimilation increments became physically consistent and persisted between updates.

Collectively, all our findings establish that energy-aware hybrid models, 
when combined with ensemble-based DA, offer a powerful and computationally efficient 
solution for high-fidelity simulations and predictions in oceanic applications. 
The synergy between the classical physics-based modeling and data-driven approach, 
together with region-based observational strategies, paves the way for new capabilities 
in both research and operational forecasting.

Despite these advances, several important questions remain open and point to 
promising avenues for future study:\\ 
$\bullet$ How robust are these methods 
when applied to more complex, fully coupled ocean-atmosphere models or under 
real-world data constraints, including missing or noisy observations? \\

$\bullet$ What is the optimal design of observational networks in more heterogeneous and 
dynamically evolving flow regimes?\\ 

$\bullet$ How can the assimilation framework be adapted to incorporate other sources of information, 
such as satellites or drifter data?\\

$\bullet$ Finally, what are the computational trade-offs and scalability challenges when 
extending this methodology to operational, global prediction systems?

\noindent
Addressing these open questions will be crucial for translating the demonstrated benefits of 
energy-aware hybrid models with ensemble DA into practical tools for next-generation 
Earth system modeling and prediction.

\acknowledgments
Igor Shevchenko thanks the Natural Environment Research Council for the support of this work through 
the project AtlantiS (P11742).
Dan Crisan thanks the European Research Council (ERC) under the European Union’s
Horizon 2020 Research and Innovation Programme for the partial support of this work through ERC, Grant Agreement No 856408.


%
%

\bibliography{refs}

%
%
%
%
%

\end{document}